\documentclass[aps,prl,showpacs,superscriptaddress,groupedaddress]{revtex4-2}

\usepackage{epsfig}\usepackage{amssymb}
\usepackage{amsmath}
\usepackage{amsfonts}
\usepackage{graphicx}
\usepackage{caption}
\usepackage{subfigure}
\usepackage{subfloat}
\usepackage{float}
\usepackage{mathrsfs}
\usepackage{multirow}
\usepackage{wrapfig}
\usepackage{hyperref}
\usepackage{dcolumn}   
\usepackage[dvipsnames]{xcolor}

\newcommand{\be}{\begin{equation}}
\newcommand{\ee}{\end{equation}}
\newcommand{\bea}{\begin{eqnarray}}
\newcommand{\eea}{\end{eqnarray}}

\newcommand{\ed}{\end{document}}

\newcommand{\bi}{\begin{itemize}}
\newcommand{\ei}{\end{itemize}}

\newcommand{\bce}{\begin{center}}
\newcommand{\ece}{\end{center}}

\usepackage{tikz}
\usetikzlibrary{calc,tikzmark} 
\newlength{\imageheight}
\begin{document}

\title{Exploring Spectral Singularities in Dirac Semimetals: The Role of Non-Hermitian Physics and Dichroism}

\author{Mustafa Sar{\i}saman}\email{mustafa.sarisaman@istanbul.edu.tr}\affiliation{National Intelligence 
Academy, Institute of Engineering and Science, Ankara, Turkey}\affiliation{Department of Physics, 
Istanbul University, 34134, Vezneciler, Istanbul, Turkey}
\author{Murat Tas}\email{murat.tas@gtu.edu.tr}\affiliation{Department of Physics, Gebze Technical 
University, 41400 Kocaeli, Turkey}
\author{Enes Talha K{\i}rca}\email{taleskirca@gmail.com}\affiliation{Department of Physics, Istanbul 
University, 34134, Vezneciler, Istanbul, Turkey}

\begin{abstract}
In this study, motivated by recent advancements in non-Hermitian physics, we explore new characteristics 
of Dirac semimetals (DSMs) using the spectral singularities by means of scattering techniques, with the 
goal of uncovering additional unique properties. To achieve this, we investigate how the axion texture 
of a DSM affects its topological properties by analyzing its interaction with electromagnetic waves. We 
examine the transverse electric (TE) mode configuration, where the magneto-electric effect induces a 
dichroic property in these materials. This behavior is particularly interesting and commonly seen in 
potential DSM candidates. Consequently, we report for the first time that a dichroic DSM generates 12 
unique topological laser types. We discover that surface currents are generated by topological terms on 
the surface of the DSM slab. Furthermore, we examine how the $\theta$ term associated with axions in 
topological materials contributes to these topological properties. Our study reveals distinct topological 
role of the $\theta$ term more clearly than ever before. Our results confirm that the topological 
properties of DSMs with a single Dirac cone remain stable under external influences and that a 
topologically robust DSM laser can be developed accordingly.
\end{abstract}

\pacs{03.50.De, 03.65.−w, 03.65.Nk, 42.25.Bs, 42.25.Gy, 42.55.−f, 71.55.Ak, 78.20.−e, 81.05.Bx}

\maketitle

\section{Introduction}

It is fascinating to discover that certain advanced mathematical concepts, such as topology, have direct 
physical counterparts with practical applications \cite{topphys1, topphys2}. This realization has opened 
up new avenues for research, particularly in understanding topological phases of matter and their 
integration into material science, giving rise to novel application areas \cite{topphase1, topphase2, 
topphase3}. Among various topological materials that have been identified, some of the most prominent 
include electronic and photonic topological insulators, topological superconductors, and Dirac and Weyl 
semimetals and metals \cite{topmats1, topmats2, topmats3, topmats4, topmats5, topmats6, topmats7, 
topmats8, topmats9, topmats10, topmats11, topmats12, topmats13}. One class of materials that has garnered 
significant attention recently is the topological Dirac semimetals (DSMs), where the valence and 
conduction bands touch at specific points in momentum space, known as Dirac points \cite{DSM1, DSM2, 
DSM3, DSM4, DSM5, DSM6, DSM7, DSM8, DSM9, DSM10, DSM11, DSM12}. What distinguishes these materials as 
``topological" is the unique arrangement of degenerate Weyl nodes that act as an axion term, imparting a 
topological nature to the system \cite{axion1, axion2, axion3}. 

While many properties of DSMs have already been explored, there remains a gap in understanding of their 
optical interactions and topological implications of these interactions. This study aims to address this 
gap and provide further insights into the optical and topological response of DSM systems \cite{diracopt1, diracopt2, diracopt3, diracopt4, diracopt5, diracopt6, 
diracopt7, diracopt8, diracopt9, diracopt10, diracopt11, diracopt12, diracopt13, diracopt14, diracopt15, 
diracopt16, diracopt17}.

In ever-evolving landscape of condensed matter physics, DSMs have emerged as a groundbreaking class of 
materials captivating scientists with their unique electronic properties and potential applications. 
These materials are characterized by the presence of Dirac cones in their electronic band structures. 
This intriguing band structure imparts DSMs with remarkable characteristics such as high mobility of 
charge carriers, low effective mass, exotic topological states and unusual response to external fields. 
DSMs stand at the intersection of several fundamental concepts in physics. They exhibit linear dispersion 
relation near the Fermi level, akin to that observed in two-dimensional graphene. These three-dimensional 
(3D) analogs of graphene not only expand our understanding of topological phases of matter but also open 
avenues for practical applications. Indeed, DSMs exhibit a range of unique properties that make them 
suitable for various cutting-edge applications. Their potential spans multiple advanced technologies, 
including quantum computing, electronics, spintronics, thermoelectrics, photonics \cite{Na3Bi6, Na3Bi7}. 

In this study, we explore the fundamental principles underlying the non-Hermitian aspects of DSMs, 
focusing on their electronic structure and the role played by the Dirac cones via the non-Hermitian 
scattering formalism. We examine DSM lasers by means of their associated spectral singularities, review 
recent developments in characterization of such materials in this context, discuss their theoretical 
models, and highlight their potential impact on future technological innovations. By integrating 
theoretical insights with experimental breakthroughs, this article aims to provide an overview of DSMs 
and their place in the cutting-edge field of materials science within the context of non-Hermitian 
physics. In the realm of advanced photonics and materials science, the fusion of DSMs with non-Hermitian 
physics represents a frontier of unprecedented potential. This innovative intersection promises not only 
deepens our understanding of quantum materials but also paves the way for revolutionary advancements in 
laser technology. When subjected to non-Hermitian principles-typically applied to open systems with gain 
and/or loss-these materials exhibit novel behaviors that are both theoretically intriguing and 
technologically advantageous.

The non-Hermitian physics explores systems where energy and other observables can exhibit non-conservative 
dynamics, often due to the presence of gain and/or loss. In such systems, conventional notions of quantum 
mechanics are modified, leading to novel effects such as exceptional points, unidirectional light 
propagation, and enhanced lasing performance \cite{bender, ijgmmp-2010, longhi4, longhi3, nonhermit1, 
nonhermit2, nonhermit3, nonhermit4, nonhermit5, nonhermit6, nonhermit7, nonhermit8, nonhermit9, 
nonhermit10, nonhermit11, nonhermit12, nonhermit13, grapheneenergy, Mandal, Mandal2}. At exceptional points, although the 
system may have real eigenvalues, the eigenstates may coalesce. In the case of scattering, the spectral 
singularities of any optical system in this picture correspond to states of divergent reflection and 
transmission amplitudes for the real $k$ values of the physical system \cite{p123, prl-2009, CPA, 
lastpaper, pra-2011a, pra-2012a, ramezani, jin1, xu1}. This causes the zero-width resonance and the laser threshold state to 
occur, as it generally produces purely outgoing waves \cite{prl-2009}. This is a natural consequence of 
non-Hermitian physics, unlike traditional lasers. In recent years, very essential studies have been 
carried out for understanding many unknown aspects of new phenomena and realities with non-Hermitian 
physics, and numerous studies have recently been devoted to exploring these phenomena\cite{nonhermit1, nonhermit2, 
nonhermit3, nonhermit4, nonhermit5, nonhermit6, nonhermit7, nonhermit8, nonhermit9, nonhermit10, 
nonhermit11, nonhermit12, nonhermit13}. Non-Hermitian physics thus plays a crucial role in understanding 
exotic properties of topological materials \cite{sarisaman1, hamed2020, sarisaman2019, sarisaman20192}. 
This idea constitutes the main motivation of our work. Examining topological systems with non-Hermitian 
physics is a very interesting and remarkable approach. By integrating DSMs into these non-Hermitian 
frameworks, researchers are uncovering pathways to engineer lasers with unprecedented efficiency, 
tunability, and robustness \cite{jin2, ota}.

Given the growing interest in this field and the fact that our topological material of focus has an 
optically active structure, we will examine how it interacts with electromagnetic waves. Recently, 
discovery of dichroism effects in DSMs has shown how important it is to study these interactions. Our 
goal here is to investigate and extract the dichroism effect in these materials. Just as Kerr and Faraday 
rotations in Weyl semimetals lead to an increase in the system size, dichroism leads to a similar result 
in a DSM, such that one encounters an additional computational challenge in the structure 
\cite{sarisaman1, ntws}. However, this challenge also reveals deeper insights into the system. Through 
our investigations, we uncover previously unknown aspects of DSMs by exploring these complexities. To 
address this task, we designed our system so that the dichroism effect results in a $4 \times 4$ transfer 
matrix, leading to 12 distinct lasing configurations, some of which have topologically robust features. 
We aim to explore topological effects in our system by generating waves in the TE mode, which will allow 
us to study the topological characteristics of DSMs. It is well-established that topological properties 
of DSMs are governed by so-called the $\theta$ term \cite{axion2, theta1}, which is in our case is simply 
$\pi$.

To understand how the $\theta$ term determines topological properties of DSMs, we analyze scattering 
behavior of them, identify their spectral singularities, and investigate influence of $\theta$ term on 
these singularities. Spectral singularities are the points where continuous spectrum of a system displays 
exceptional characteristics \cite{naimark, naimark-1, p123}. Thus, interaction of a DSM with 
electromagnetic waves can be viewed as a non-Hermitian scattering problem in electromagnetic theory 
\cite{sarisaman1}.

Our work proceeds as follows. First, we calculate the transfer matrix through boundary conditions by 
solving Maxwell equations with an axion term specific to DSM for the TE mode configuration. This transfer 
matrix allows us to calculate the spectral singularities. By calculating the spectral singularities in 
this way, we determine effect of the $\theta$ term on the spectral singularities of Na$_3$Bi, which has 
been experimentally proven to be a DSM \cite{diracopt4, Na3Bi2, Na3Bi3, Na3Bi4, Na3Bi5}. In our 
investigations, we obtain novel results, such as dichroism in DSMs is an absolute effect and it gives 
rise to 12 different topological laser types. Accordingly, presence of the $\theta$ term significantly 
reduces the gain value in the system. It is manifestly shown that the gain is topologically quantized by 
degenerating spectral singularity points in the system. This result is very important and has been 
noticed for the first time. We finally find out that an induced current presents on the surfaces of a DSM. 
Notable results of this study show that 12 different topological laser types can be created due to the 
dichroism effect in a DSM, and under what conditions these lasers can exist.

\section{Presence of The $\theta$ term and Electrodynamics in a Planar DSM}
\label{S1}

Before proceeding with the formal derivation, it is useful to briefly clarify the theoretical framework adopted in this work. Our analysis is based on the effective electromagnetic response of Dirac semimetals rather than a microscopic band-Hamiltonian formulation. At low energies, the electronic degrees of freedom in topological semimetals can be integrated out, leading to an effective field theory where the electromagnetic response is governed by Maxwell equations supplemented by an axion term characterized by the parameter $\theta$. This approach, commonly referred to as axion electrodynamics, provides a well-established description of the topological magneto-electric response in Dirac and Weyl semimetals.

Within this effective description, the optical properties of the system can be investigated through the scattering formalism of electromagnetic waves interacting with the material. In such open systems the presence of gain or loss naturally leads to a non-Hermitian framework, where spectral singularities appear as real-frequency poles of the scattering amplitudes. These singularities correspond to zero-width resonances and are known to represent the laser threshold condition in non-Hermitian optical systems. While exceptional points are typically discussed in the context of non-Hermitian Hamiltonians where eigenvalues and eigenvectors coalesce, spectral singularities constitute their scattering-theoretic counterpart. In the present work we therefore focus on the emergence of spectral singularities within the transfer-matrix formulation of the DSM slab system and analyze their physical consequences for the generation of topological laser modes.

Understanding wave propagation in a DSM environment requires grasping the role of $\theta$ term arising 
from material properties. The $\theta$ term is known as the magneto-electric polarizability, and is 
related to the Berry phase and Chern number. Its mission in electrodynamic interactions is accounted for 
axion electrodynamics. In Table \ref{table1}, we provide its values in different materials.

\begin{table}[!htb]
\renewcommand{\arraystretch}{1.5}
\caption{The $\theta$ term for various material types. Here $b^{\mu}=(b_0, \vec{b})$,
$x^{\mu}=(t, \vec{x})$, and the Minkowski metric is assumed to be $\eta=\textrm{diag} (-1, +1, +1, +1)$.}
\begin{tabular}{c|c} \hline
 $\theta$           & Material Type            \\ \hline\hline
 0                  & ~~ Ordinary insulators   \\ \hline
 $\pi$              & ~Time-reversal symmetric topological insulators \\ \hline
 $2b_{\mu}x^{\mu}$~ & Weyl semimetals          \\ \hline
 $\pi$              & Dirac semimetals         \\ \hline
\end{tabular}
\renewcommand{\arraystretch}{1.5}
\label{table1}
\end{table}

\begin{figure}[!hbt]
\includegraphics[width=5cm]{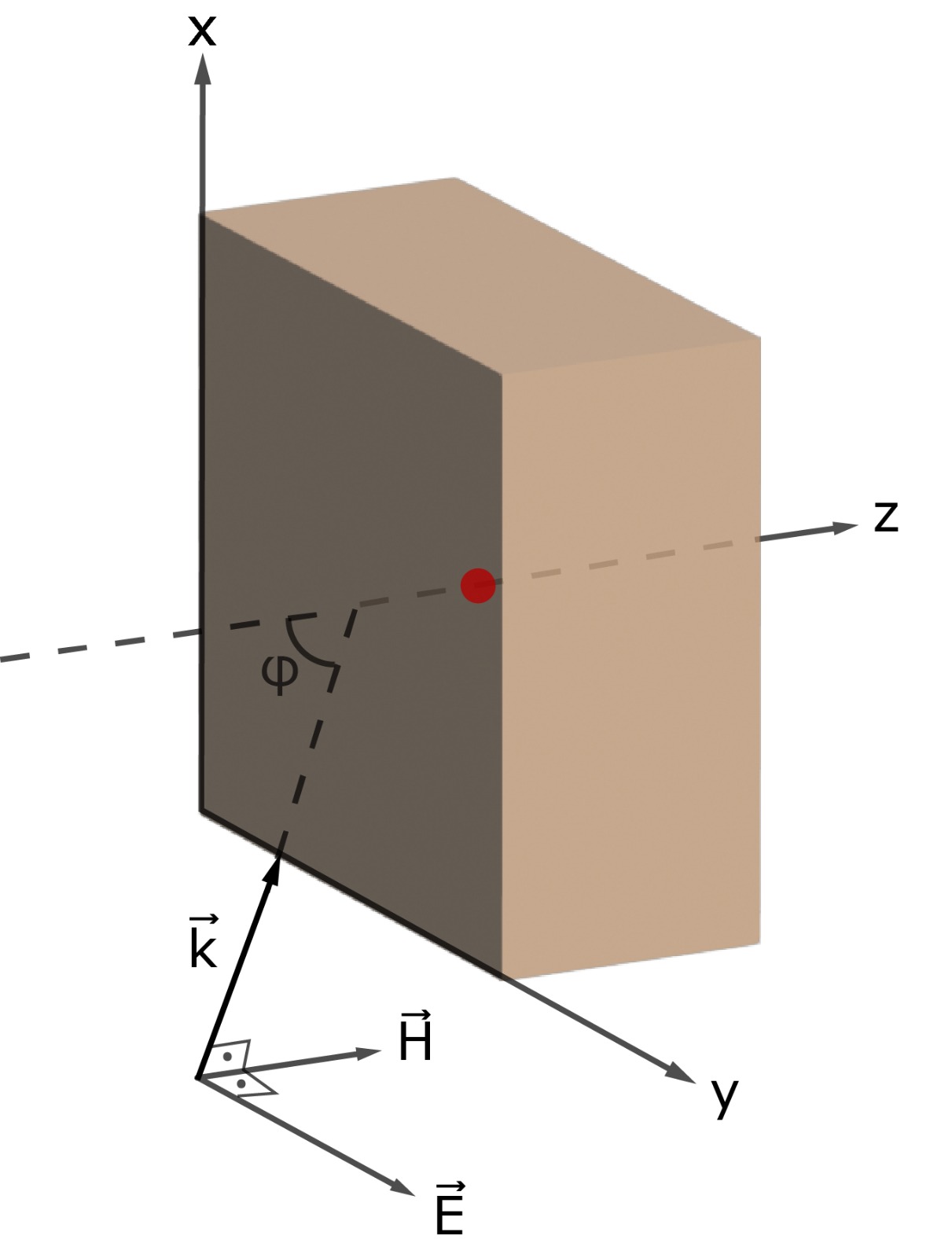}
\includegraphics[width=5cm]{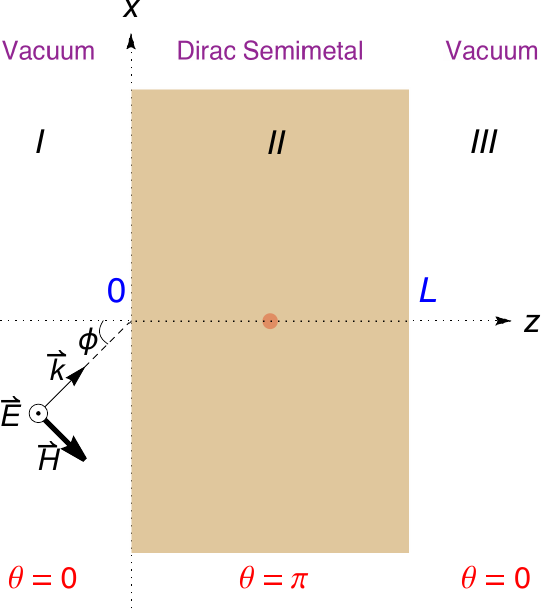}~~~~~~~
\includegraphics[width=5cm]{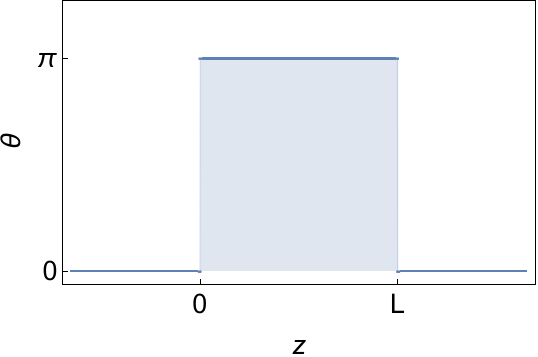}
\caption{TE mode configuration for the interaction of electromagnetic wave incident by an angle $\phi$ 
measured from the normal to the surface of DSM slab. Left panel displays the 3D configuration while 
middle panel shows the top view. Red dot identifies the single Weyl node. $\theta$ terms corresponding 
to each medium are demonstrated on the right panel. Colored region specifies the DSM.}
\label{fig1}
\end{figure}

We consider an optically active linear, homogeneous 3D slab of DSM with thickness $L$ aligned along the 
$z$-axis as shown in Fig.~\ref{fig1}. Although temperature, disorder, and impurities can affect the 
physical and topological properties of the system, we will simplify our analysis by focusing solely on 
the effect of the $\theta$ term. Therefore, we consider a material that is linear, homogeneous, and 
unaffected by temperature. The $\theta$ term, which survives both inside and at the boundaries of the 
DSM, can be expressed as a function of $z$ in the form $\theta(z) = \pi\,\Theta(z)\,\Theta (L-z)$, 
where $\Theta(z)$ is the Heaviside step function defined as
\begin{align}\label{eq18}
	\Theta(z) := \begin{cases}
                   0, \qquad z < 0  \\
		   1, \qquad z \geq 0.
	\end{cases}
\end{align}
This actually demonstrates the semimetallic nature inherent in these materials. Complex refractive index 
within the slab is considered uniform across the region $z \in [0, L]$ between its end faces. Observe 
that the 3D material extends indefinitely in the $x$ and $y$ directions. Interaction of this slab system 
with electromagnetic waves is crucial for understanding its topological and magneto-electric properties, 
as well as its applications in quantum device technologies. Topological aspects arise from the single 
placement of Weyl node, which determine the conductive character on the faces of the slab. In our set up, 
the nodes are located along the $z$-axis and a constant $\theta$ term appears inside the slab. 

Maxwell equations incorporating additional topological terms associated with the Weyl node for our slab 
configuration can be written as \cite{sarisaman1, axion3}
\begin{align}
    \label{1}   &\vec{\nabla}\cdot\vec{D}_{\ell}  = \rho\,, \\
    \label{2}   &\vec{\nabla}\cdot\vec{B}_{\ell}  = 0\,, \\
    \label{eq3} &\vec{\nabla}\times\vec{E}_{\ell} =-\partial_t\vec{B}_{\ell}\,, \\
    \label{eq4} &\vec{\nabla}\times\vec{H}_{\ell} = \vec{J} +\partial_t\vec{D}_{\ell}\,.
    \end{align}    
In these equations, $\vec{D}_{\ell}$, $\vec{B}_{\ell}$, $\vec{E}_{\ell}$ and $\vec{H}_{\ell}$ denote 
linear expressions assumed by the electromagnetic fields in the presence of $\theta$ term. $\rho$ and 
$\vec{J}$ denote, respectively, the charge and current densities, which depend on position and time. 
These expressions are free and axion-induced expressions, and can be written as: 
$\rho := \rho_f + \rho_{\theta}$ and $\vec{J} := \vec{J}_f + \vec{J}_{\theta}$. 
Eqs.~(\ref{1})-(\ref{eq4}) are Maxwell equations in a DSM environment with a $\theta$ term. Two 
different approaches can be followed to obtain these equations. The first of these is to find equations 
of motion by writing the action expression that specifies interaction between the electromagnetic 
fields and the $\theta$ term. The other one is to show interaction of the $\theta$ term with relevant 
fields by direct reflection. These are shown in the Appendix.

\subsection{Continuity Equation and Axion-Induced Surface Currents} \label{S23} 

Quantities $\rho$ and $\vec{J}$ in Eqs.~(\ref{1}) and (\ref{eq4}) satisfy the continuity equation as a 
natural consequence of Maxwell equations:
\begin{equation}
\vec{\nabla} \cdot \vec{J} + \partial_t \rho=0 \label{continuityeqn}
\end{equation}
These quantities may be decomposed into free and induced-axion terms, such that two distinct continuity 
equations are obtained as follows:
\begin{align}
&\vec{\nabla} \cdot \vec{J}_f = 0 \label{continuityeqnfree}\\
&\vec{\nabla} \cdot \vec{J}_{\theta} + \partial_t \rho_{\theta}=0 \label{continuityeqnaxion}
\end{align}
Note that here the free charge density is taken as zero. However, due to the semimetallic character of 
the material, the free current density is different from zero. In this case, divergence of the free 
current density is zero. The free current density can be related to the conductivity tensor of the 
material. In this case,
\begin{equation}
\vec{J}_f = \sigma \vec{E}_{\ell} \qquad \Rightarrow \qquad \left[J_f\right]_{\alpha} = 
\sigma_{\alpha\beta} E_{\beta}\,. \label{currentdensity}
\end{equation}
Here $\sigma$ can be briefly represented as
\begin{equation}
\sigma = \begin{pmatrix} \sigma_{yy} & \sigma_{yz}\\ \sigma_{zy} & \sigma_{zz} \end{pmatrix}
\end{equation}
It can be seen that $\sigma_{zz}$ component vanishes due to material properties.

\subsection{Scattering Solutions and Dichroism Effect in a Planar Dirac Semimetal} \label{S4}  

Since the material considered here contains an axionic content, it is inevitable that exotic surface 
effects will appear when exposed to an external electromagnetic wave. The axionic contribution of the 
material is considered along the $z$-axis only. Thus, if the electromagnetic wave incident from outside 
has a magnetic field in this direction, a rotation in the polarization direction of the wave is expected. 
A judicious choice should be made and polarization direction of the electromagnetic wave incident from 
outside should be sent in a way that it would interact with the axionic term. This corresponds to the TE 
mode configuration in our study. Hence, a wave polarized along the $y$-axis and making an angle of 
$\phi$ with the surface normal was sent to the DSM medium from outside, and result of its interaction 
with the DSM is aimed to be examined within the scope of scattering theory. This choice ensures that the electromagnetic wave couples to the axionic magneto-electric response of the material, which is responsible for the emergence of dichroic behavior. Propagation of the wave in a 
material is controlled by the solutions of Maxwell equations. Accordingly, by taking the curl (i.e. 
$\vec{\nabla} \times$) of Eq.~(\ref{eq3}), and then using Eq.~(\ref{eq4}), 3D Helmholtz equation is 
obtained as
\begin{equation}
\nabla^2 \vec{E} + \varepsilon_{0} \omega^2 \mathbf{n}^2 \vec{E} + i \omega \mu_0 \vec{J} = 0 \,.
\label{3dhelmholtz}
\end{equation}
Here, $\mathbf{n}$ represents the complex refractive index of the DSM. For an active material medium 
with a gain or loss content, the refractive index can be decomposed into its real and imaginary parts as  
\begin{equation}
\mathbf{n} = \eta + i\kappa\,. \label{refractiveindex}
\end{equation}
For various materials, including DSMs, it is in general safe to assume that $|\kappa| \ll \eta$, where 
the parameter $\kappa$ essentially dictates whether the medium exhibits gain or loss. Specifically, when 
$\kappa < 0$, the medium functions as a gain medium, while for $\kappa > 0$ the medium behaves as an 
absorbing or loss medium.

When a wave is incident to the DSM medium from outside with TE mode in $y$-direction, some effects will 
occur when the wave hits the surface due to the electric and magnetic properties of the material. 
Boundary conditions of the material indicate that the incident wave will not create a (Kerr/Faraday) 
rotation effect in $x$-direction. In this case, it is understood that such types of materials cannot 
cause a Faraday or Kerr type rotations. However, since the axionic content increases dimension of the 
electromagnetic wave scattered from the material, the only remaining option is that the polarization 
direction of the electromagnetic wave in the material will rotate towards the $z$-axis. This is known as 
the dichroism effect and indicates rotation of the wave along the material direction. This effect has 
actually been observed in such materials, but it is a non-generalized effect that is presented for the 
first time in this study. The results of this study show that the DSM medium is a dichroic medium. Thus, 
an electromagnetic wave polarized in the $y$-direction will move in the material such a way that it will 
be polarized in the $y-z$ plane after the scattering event. In this case, the electric field solution 
after the scattering can be written as 
\begin{equation}
\vec{E}(x, z) = E_y(x, z)\,\hat{e}_y + E_z(x, z)\,\hat{e}_z \,.
\end{equation}
This expression is inserted into Eq.~(\ref{3dhelmholtz}) to obtain the scattering solutions inside and 
outside of the DSM. It should be noted that current $\vec{J}$ is zero outside the material. To understand 
characteristics of these solutions inside the material, values of $\vec{J}$ are important. As is clearly 
seen, the free current density that forms $\vec{J}_f$ exists inside the material due to the dichroism 
effect. However, induced current $\vec{J}_{\theta}$ originating from the axionic term appears only on the 
surfaces. The reason for this is the presence of Dirac delta functions in the $\vec{J}_{\theta}$ 
expression. In this case, free current densities inside the material can be calculated using 
Eq.~(\ref{currentdensity}). Another point to note is that the dichroism effect only causes existence of 
the conductivity components $\sigma_{yz}$ and $\sigma_{zy}$. In this case, it can be seen that the 
Helmholtz equation (\ref{3dhelmholtz}) can be reduced to the following form,
\begin{equation}
\nabla^2 E_{\alpha} + \varepsilon_0 \omega^2 \mathbf{n}^2 E_{\alpha} + i \omega\mu_0\sigma_{\alpha\beta} 
E_{\beta} = 0 \quad \Rightarrow  \quad  \left[\nabla^2 \delta_{\alpha\beta} + \varepsilon_0 \omega^2 
\mathbf{n}^2 \delta_{\alpha\beta} + i\omega\mu_0\sigma_{\alpha\beta}\right] E_{\beta} = 0 \,.
\end{equation}
Here the subscripts $\alpha$ and $\beta$ represent $y$ and $z$ components of the electric field. 
Describing these equations in terms of components yields the following coupled relations:
\begin{align}
\nabla^2 E_{y}+ \varepsilon_0 \omega^2 \mathbf{n}^2 E_{y}+ i\omega\mu_0\sigma_{yz} E_{z}=0 \label{eq326}\\
\nabla^2 E_{z}+ \varepsilon_0 \omega^2 \mathbf{n}^2 E_{z}+ i\omega\mu_0\sigma_{zy} E_{y}=0 \label{eq327}
\end{align}
Note that $\sigma_{yz}=\sigma_{zy}$ due to the symmetric nature of $\sigma$. One can solve these by 
multiplying Eq.~(\ref{eq326}) with $i$, and then add and subtract it from Eq.~(\ref{eq327}), yielding the 
following pair of equations:
\begin{align}
\left(\nabla^2+\varepsilon_0\omega^2\mathbf{n}^2 \right)\Phi_{\pm} \mp \omega\mu_0\sigma \Phi_{\mp} = 0\,.
\end{align}
Here, $\Phi_{\pm}:= E_z \pm i E_y$ and $\sigma:= \sigma_{yz} = \sigma_{zy}$. Writing this equation 
explicitly results in coupled new expressions,
\begin{align}
\left(\nabla^2 + \varepsilon_0 \omega^2 \mathbf{n}^2\right) \Phi_{+} - 
\omega\mu_0\sigma\Phi_{-} = 0 \label{eq329} \\
\left(\nabla^2 + \varepsilon_0 \omega^2 \mathbf{n}^2\right) \Phi_{-} + 
\omega\mu_0\sigma\Phi_{+} = 0 \label{eq330}
\end{align}
To solve these coupled relations, Eq.~(\ref{eq330}) is multiplied by $i$, then added to and subtracted 
from Eq.~(\ref{eq329}). These operations yield the following decoupled equations,
\begin{align}
\left(\nabla^2 + \varepsilon_0 \omega^2 \mathbf{n}^2 \right) \Psi_{\pm} \pm 
i \omega\mu_0\sigma \Psi_{\pm} = 0 \,, \label{eq331}
\end{align}
where $\Psi_{\pm} := \Phi_{+} \pm i\,\Phi_{-}$. These solutions in terms of the electric field components 
can be calculated easily as
\begin{equation}
\Psi_{\pm} := (1 \pm i)(E_z \pm E_y)\,.
\end{equation}
Thus, components $E_y$ and $E_z$ can be calculated in terms of new fields $\Psi_{+}$ and $\Psi_{-}$ as 
\begin{align}
E_y = \frac{1}{4}\left[(1-i) \Psi_{+} - (1+i)\Psi_{-}\right] \,, \label{eq333} \\
E_z = \frac{1}{4}\left[(1-i) \Psi_{+} + (1+i)\Psi_{-}\right] \,. \label{eq334}
\end{align}
It should be noted that arguments of these quantities depend on $x$ and $z$. As a result, solutions of 
the decoupled Helmholtz equation (\ref{eq331}) can now be easily found. For this purpose, the arguments 
can be separated as $\Psi_{\pm}(x, z) = e^{ik_x x} \Psi_{\pm}(z)$. When these expressions are 
substituted into Eq.~(\ref{eq331}), the 1D Helmholtz equation with respect to $z$ is obtained 
\begin{equation}
\Psi''_{\pm}(z) + k_z^2 \mathbf{\tilde{n}}_{\pm}^2 \Psi_{\pm}(z) = 0\,.  \label{eq335}
\end{equation}
In this equation, the prime($\prime$) symbol indicates the derivative with respect to $z$, 
$k_z = k \cos \phi$ is the $z$-component of the wave-vector $\vec k$, $\mathbf{\tilde{n}}_{\pm}$ is the 
frequency dependent birefringence index, which appears due to the conductivity of the material, and it 
is defined as
\begin{equation}
\mathbf{\tilde{n}}_{\pm} := \sqrt{\mathbf{\tilde{n}}^2 \pm \frac{i \mu_o \omega \sigma}{k_z^2}}, 
\qquad \qquad \mathbf{\tilde{n}} := \frac{\sqrt{\mathbf{n}^2-\sin^2\phi}}{\cos\phi}\,.
\end{equation}
Notice that Eq.~(\ref{eq335}) gives rise to the following Schr{\"{o}}dinger equation with a constant 
complex potential $V_{\pm}(z) := k_z^2 (1-\mathbf{\tilde{n}}_{\pm}^2)$
\begin{equation}
-\Psi''_{\pm}(z) + V_{\pm}(z) \Psi_{\pm}(z) = k_z^2 \Psi_{\pm}(z)\,.  \label{schrodingereq}
\end{equation}

As a natural outcome of this analysis, materials exhibiting a DSM phase have two distinct refractive 
indices (a birefringence effect), with each refractive index corresponding to its own unique 
Schr{\"{o}}dinger equation. This indicates a non-Hermitian state with real energy values, corresponding 
to the scattering state we described earlier. Here $\mathbf{\tilde{n}}$ represents the effective 
refractive index of the material, and is caused by the wave arriving at a certain angle. Thus, solution 
to the Helmholtz equation, as detailed in Eq.~(\ref{eq335}), is expressed as follows:
\begin{align}
\Psi_{\pm}(z) := \begin{cases}
A_{1\pm}\,e^{ik_z z} +  B_{1\pm}\, e^{-ik_z z}, \hskip 1.3cm z < 0\\
A_{2\pm}\,e^{ik_z\mathbf{\tilde{n}}_{\pm}z} + B_{2\pm}\, e^{-ik_z\mathbf{\tilde{n}}_{\pm}z}, 
\hskip 0.56cm 0 < z < L\\
A_{3\pm}\,e^{ik_zz} +  B_{3\pm}\, e^{-ik_zz}, \hskip 1.3cm z > L.
	\end{cases} \label{Fpm}
\end{align}
These results can be substituted into Eqs.~(\ref{eq333}) and (\ref{eq334}) in order to obtain electric 
field components $E_y$ and $E_z$. Similarly, to find magnetic field $\vec{B}$, values of $E_y$ and $E_z$ 
are substituted into Eq.~(\ref{eq317}). As a result, the field $\vec{B}(x, z)$ is expressed as:
\begin{align}
\vec{B} = -\frac{i}{\omega} \left(\vec{\nabla} \times \vec{E}\right) = \frac{i}{\omega} 
\left[ \hat{e}_x\,\partial_{z} E_y + \hat{e}_y\,\partial_x E_z - \hat{e}_z\,\partial_x E_y  \right] \,.
\end{align}
But since $E_{y}(x, z) = e^{ik_x x} \mathcal{E}_y(z)$, we find
\begin{equation}
\partial_x E_y = i k_x E_y \,. \notag
\end{equation}
As a result, these quantities can be made independent of $x$, and $\vec{\mathcal{B}}(z)$ is obtained as
\begin{align}
\vec{\mathcal{B}}(z) = \frac{i}{\omega} \left[ \partial_z \mathcal{E}_y \hat{e}_x + 
i k_x \mathcal{E}_z \hat{e}_y -  i k_x \mathcal{E}_y \hat{e}_z \right]\,.
\end{align}
It will be seen that $\partial_z \mathcal{E}_y$ is found as follows
\begin{equation}
\partial_z \mathcal{E}_y = \frac{(1-i)}{4}\,\partial_z \Psi_{+} - \frac{(1 + i)}{4}\,\partial_z \Psi_{-}
\end{equation}
with 
\begin{equation}
\partial_z \Psi_{\pm} := i k_z \mathbf{\tilde{N}}_{\pm} \mathcal{F}_{\pm}^{-}, \quad 
\mathbf{\tilde{N}}_{\pm} := \begin{bmatrix} 1 \\ \mathbf{\tilde{n}}_{\pm}\\ 1 \end{bmatrix}, \quad 
\mathcal{F}_{a}^{\ell} (z) := \begin{bmatrix} A_{1 a}\,e^{ik_z z} + \ell B_{1 a}\,e^{-ik_z z} \\ 
A_{2 a}\,e^{ik_z \mathbf{\tilde{n}}_{a} z} + \ell B_{2 a}\,e^{-ik_z \mathbf{\tilde{n}}_{a} z} \\ 
A_{3 a}\,e^{ik_z z} + \ell B_{3 a}\,e^{-ik_z z} \end{bmatrix}\,, \label{eq341}
\end{equation}
where the subscripts $a$ and $\ell$ take the signs $+ / -$. Consequently, with these new definitions, the 
electric and magnetic field components can be determined as in Table~\ref{t1}. Likewise, one can obtain 
the $\vec{H}$ and $\vec{D}$ fields using definitions: $\vec{H} = \vec{B} / \mu_0$ and 
$\vec{D} = \varepsilon\,\vec{E}$; their components are given in Table~\ref{t1}. Symbol $\mathbf{N}$, 
defined in Table~\ref{t1}, represents refractive index of the entire space and is given by:
\begin{equation}
\mathbf{N} := \begin{bmatrix} 1 \\ \mathbf{n} \\ 1 \end{bmatrix} \label{allrefr}
\end{equation}
Notice that $\mathbf{N}$ and $\mathbf{\tilde{N}}$ can be related to each other by 
$\mathbf{\tilde{N}} = \sqrt{\mathbf{N}^2 -\sin^2\phi} / \cos\phi$, where $\mathbf{\tilde{N}}$ represents 
effective refractive index of the entire space.

\begin{table}[!hbt]
\caption{Components of the fields $\vec{\textbf{Z}} \in \left\{ \vec{E}, \vec{D}, \vec{B}, 
\vec{H} \right\}$ inside and outside of the DSM slab. Here 
$\vec{\textbf{Z}} (x, z) := \frac{(1-i)\,e^{ik_x x}}{4} \vec{\mathcal{Z}} (z)$ and 
$\vec{\mathcal{Z}} \in \left\{\vec{\mathcal{E}}, \vec{\mathcal{D}}, \vec{\mathcal{B}}, 
\vec{\mathcal{H}}\right\}$. Quantities $\mathcal{F}_{a}^{\ell}(z)$ are specified in Eq.~\ref{eq341},and 
the refractive index of whole space is identified in Eq.~\ref{allrefr}.} \label{t1}
\scalebox{0.95}{\begin{tabular}{|c|c|c|c|}
\hline
\textbf{$\vec{\mathcal{E}} (z)$}& \textbf{$\vec{\mathcal{D}} (z)$}& \textbf{$\vec{\mathcal{B}}(z)$} & 
\textbf{$\vec{\mathcal{H}}(z)$}\\ \hline\hline \
$\mathcal{E}_x = 0$ & $\mathcal{D}_x = 0$ & $\mathcal{B}_x = -\frac{\cos\phi}{c} \left[\mathbf{
\tilde{N}}_{+}\, \mathcal{F}_{+}^{-}(z) - i \,\mathbf{\tilde{N}}_{-}\, \mathcal{F}_{-}^{-}(z) \right]$ & 
$\mathcal{H}_x = -\frac{\cos\phi}{Z_0} \left[\mathbf{\tilde{N}}_{+}\, \mathcal{F}_{+}^{-}(z) - i 
\,\mathbf{\tilde{N}}_{-}\, \mathcal{F}_{-}^{-}(z) \right]$\\ 
$\mathcal{E}_y = \left[\mathcal{F}_{+}^{+}(z) - i \, \mathcal{F}_{-}^{+}(z)\right]$ & $\mathcal{D}_y = 
\varepsilon_0 \mathbf{N}^2 \left[\mathcal{F}_{+}^{+}(z) - i\mathcal{F}_{-}^{+}(z)  \right] $ & 
$\mathcal{B}_y = -\frac{\sin\phi}{c} \left[\,\mathcal{F}_{+}^{+}(z) + i\, \mathcal{F}_{-}^{+}(z) \right]$ & 
$\mathcal{H}_y = -\frac{\sin\phi}{Z_0} \left[\,\mathcal{F}_{+}^{+}(z) + i\, \mathcal{F}_{-}^{+}(z) \right]$\\
$\mathcal{E}_z = \left[ \mathcal{F}_{+}^{+}(z) + i \, \mathcal{F}_{-}^{+}(z)\right] $ & 
$\mathcal{D}_z = \varepsilon_0 \mathbf{N}^2 \left[\mathcal{F}_{+}^{+}(z) + i \mathcal{F}_{-}^{+}(z) \right]$ 
& $\mathcal{B}_z = \frac{\sin\phi}{c} \left[\mathcal{F}_{+}^{+}(z) - i\, \mathcal{F}_{-}^{+}(z) \right]$ & 
$\mathcal{H}_z = \frac{\sin\phi}{Z_0} \left[\mathcal{F}_{+}^{+}(z) - i\, \mathcal{F}_{-}^{+}(z) \right]$\\ 
\hline
\end{tabular}}	
\end{table}

After determining the electromagnetic fields at every point in space, boundary conditions at the edges 
of the DSM can be established. Standard boundary conditions for the DSM medium are outlined in 
Table~\ref{table2}, which is presented in Appendix. Hence, the boundary conditions given in 
Table~\ref{table2} can be written to relate the amplitudes on the left and right surfaces. Accordingly, 
the following notations can be used for the fields in regions I - II and II - III, respectively:
\begin{align}
\mathbb{S}\,\vec{\mathbf{V}}_2 = \mathbb{U}_+ \vec{\mathbf{V}}_1 \qquad \mathbb{S}\,\mathbb{P}_1\,
\vec{\mathbf{V}}_2 = \mathbb{U}_-\,\mathbb{P}_2\vec{\mathbf{V}}_3  \label{eq351}
\end{align}
Here we employed the following definitions.
\begin{equation}
    \mathbb{U}_{\pm} :=
    \begin{pmatrix}
    1 \pm \frac{k_x}{\varepsilon_0 \omega} & 1 \pm \frac{k_x}{\varepsilon_0 \omega} & 
i(1 \mp \frac{k_x}{\varepsilon_0 \omega}) & i(1 \mp \frac{k_x}{\varepsilon_0 \omega}) \\
    1 & 1 & -i & -i \\
    1 \mp \frac{Z_0 \alpha}{\cos \phi} & -(1 \pm \frac{Z_0 \alpha}{\cos \phi}) & 
-i(1 \mp \frac{Z_0 \alpha}{\cos \phi}) & i(1 \pm \frac{Z_0 \alpha}{\cos \phi}) \\
    1 \mp \frac{Z_0 \sigma}{\sin \phi} & 1 \mp \frac{Z_0 \sigma}{\sin \phi} & 
i(1 \mp \frac{Z_0 \sigma}{\sin \phi}) & i(1 \mp \frac{Z_0 \sigma}{\sin \phi})
    \end{pmatrix}  \\
 \end{equation}

\begin{equation}
\vec{\mathbf{V}}_i := \begin{bmatrix} A_{i+} \\ B_{i+} \\  A_{i-} \\ B_{i-} \end{bmatrix}, \qquad
    \mathbb{S} :=
    \begin{pmatrix}
    \mathbf{n}^2 & \mathbf{n}^2 & i\,\mathbf{n}^2 & i\,\mathbf{n}^2 \\
    1 & 1 & -i & -i \\
    \mathbf{\tilde{n}}_+ & -\mathbf{\tilde{n}}_+ & -i\,\mathbf{\tilde{n}}_- & i\,\mathbf{\tilde{n}}_- \\
    1 & 1 & i & i 
    \end{pmatrix}
\end{equation}

\begin{equation}
    \mathbb{P}_1 :=
    \begin{pmatrix}
    e^{i k_z \mathbf{\tilde{n}}_+ L} & 0 & 0 & 0 \\
    0 & e^{-i k_z \mathbf{\tilde{n}}_+ L} & 0 & 0 \\
    0 & 0 & e^{i k_z \mathbf{\tilde{n}}_- L} & 0 \\
    0 & 0 & 0 & e^{-i k_z \mathbf{\tilde{n}}_- L}
    \end{pmatrix}, \quad
    \mathbb{P}_2 :=
    \begin{pmatrix}
    e^{i k_z L} & 0 & 0 & 0 \\
    0 & e^{-i k_z L} & 0 & 0 \\
    0 & 0 & e^{i k_z L} & 0 \\
    0 & 0 & 0 & e^{-i k_z L}
    \end{pmatrix} \\
\end{equation}  

\subsection{Transfer Matrix and Spectral Singularities}

Transfer matrix can be constructed using the expressions in Eq.~(\ref{eq351}), which describe relations 
in the boundary regions of the DSM medium following wave scattering. The resulting transfer matrix for 
the DSM system is given by
\begin{equation}
\vec{\mathbf{V}}_3 = \mathbb{M} \vec{\mathbf{V}}_1, \qquad \mathbb{M} = \mathbb{P}_2^{-1}\,
\mathbb{U}_-^{-1}\,\mathbb{S}\, \mathbb{P}_1 \mathbb{S}^{-1}\,\mathbb{U}_+ \label{eq354}
\end{equation}

Transfer matrix is a very important tool that contains all the basic information in a scattering system, 
which can be used to control post-scattering behavior of the system. All parameters of the system can be 
obtained from its transfer matrix. Hence, the exceptional points of DSM environment in the non-Hermitian 
physics framework can be obtained again through the transfer matrix.

Transfer matrix given by Eq.~(\ref{eq354}) contains all information about the DSM system; in this section, 
we obtain its spectral singularities using the transfer matrix. Spectral Singularities in the case of 
gain doped DSM are associated with exceptional points or non-Hermitian phases where the system exhibits 
unusual behavior of lasing threshold condition. These correspond to zero-width resonances at real energy 
eigenstates. As can be understood from our analysis, a 1D DSM becomes 2D system due to the dichroism 
property even if electromagnetic waves incident in TE mode in 1D. This property causes the transfer 
matrix to be in the form of a $4 \times 4$ matrix. Thus, exceptional points in the system can be detected 
by imposing restrictions on the components of the transfer matrix. As it is known, the exceptional points 
have bases formed by the state vectors that form the Hilbert space of the system. In a normal Hermitian 
system, these bases, which should be orthonormal, become parallel to each other and/or their eigenvalues 
become coincident, resulting in the exceptional points. Since there are 4 orthonormal state vectors 
originating from two different modes in our system, very different exceptional point configurations can 
be produced. In this study, only a special case called the spectral singularity is examined. After a 
scattering event, the spectral singularity in the system forms only wave configurations going outwards. 
For this, if different solution modes of the system are called plus $(+)$ and minus $(-)$ mode solutions, 
different spectral singularity situations can be obtained. Each distinct case will be explored separately 
in the upcoming sections, with realistic DSM examples provided.

We define the gain coefficient of the DSM system as
\begin{equation}
g := -2 k \kappa \,,
\end{equation}
where $k = 2 \pi / \lambda$ denotes the free-space wavenumber, and $\kappa$ is the imaginary part of the refractive index $\mathbf{n}$, i.e. 
$\mathbf{n} = \eta + i \kappa$. To understand how the physical parameters of our system influence the 
spectral singularity condition we investigate, as an example, Na$_3$Bi in the slab geometry. Na$_3$Bi is 
recognized as one of the first practical DSMs \cite{diracopt4, Na3Bi2, Na3Bi3, Na3Bi4, Na3Bi5}. Its 
characteristics are \cite{Na3Bi6, Na3Bi7, silfvast}: 
\begin{align} 
\eta&=1.33,~~ L=500~\textrm{nm},~~ \phi=60^{\circ}\,. \label{specifications}
\end{align}

\begin{figure}[!h]
\includegraphics[width=8cm]{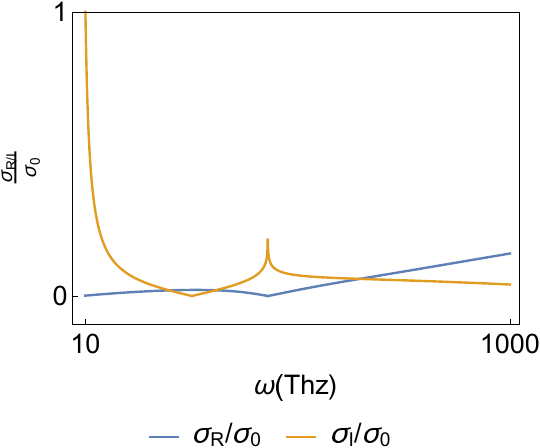}
\caption{The frequency-dependent real and imaginary parts of the conductivity for Na$_3$Bi. The 
conductivity at 10~THz frequency is indicated by $\sigma_0$.}
\label{figsigma}
\end{figure}

We notice that refractive index of DSM changes with Fermi energy, ambient temperature, and light 
frequency \cite{Na3Bi6}. We also require the conductivity of Na$_3$Bi for our computational analysis. 
According to the Kubo formalism, its conductivity in the low temperature limit can be expressed as 
$\sigma = \sigma_R + i \sigma_I$ \cite{Na3Bi6, Na3Bi7, Na3Bi8, Na3Bi9}, where the real and imaginary 
parts are approximated as follows:
\begin{align}
\sigma_R &= \frac{e^2}{\hbar} \frac{\mathtt{g} k_F}{24 \pi} \Omega \,\mathtt{G}\left(\Omega - 2\right)\,,
\notag\\
\sigma_I &= \frac{e^2}{\hbar} \frac{\mathtt{g} k_F}{24 \pi} \left[ \frac{4}{\Omega} - \Omega 
\ln \left(\frac{4\varepsilon_c^2}{|\Omega^2-4|}\right)\right]\,,
\end{align}
where $e$ is the elementary electric charge, $\mathtt{g}$ indicates the degeneracy factor, and 
$\mathtt{G}$ denotes the Riemann–Siegel theta function. The Fermi momentum is given by 
$k_F = E_F / (\hbar v_F)$, where $\hbar$ is the reduced Planck constant and $v_F$ is the Fermi velocity 
of the electrons. Other parameters are defined as $\Omega = \hbar \omega / E_F + i v_F/E_F k_F \mu$, 
where $\omega$ and $\mu$ represent the angular frequency of the incident wave and the carrier mobility, 
respectively, and $\varepsilon_c = E_c / E_F$, with $E_c$ being the cutoff energy. In our case, 
$\mathtt{g} = 40$, Fermi velocity is $v_F = 10^6$~m/s, Fermi energy level is $E_F = 0.9$~eV and 
$E_c = 3$~eV. We plot the normalized conductivity of Na$_3$Bi for these parameters in 
Fig.~\ref{figsigma}.

\subsection{Plus-Mode Spectral Singularity Configuration: Plus-Mode Topological DSM Laser}

\begin{figure}[!hbt]
\includegraphics[width=5cm]{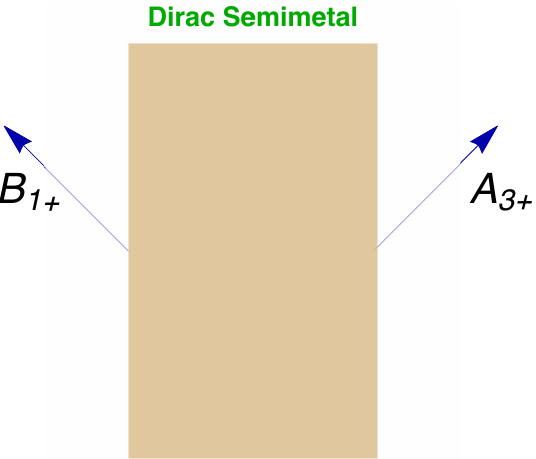}~~
\includegraphics[width=5cm]{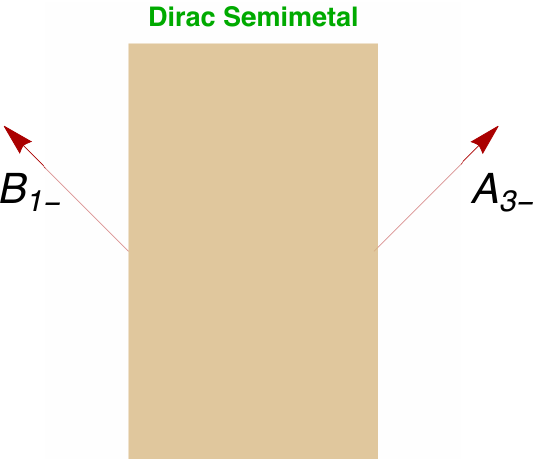}~
\includegraphics[width=5cm]{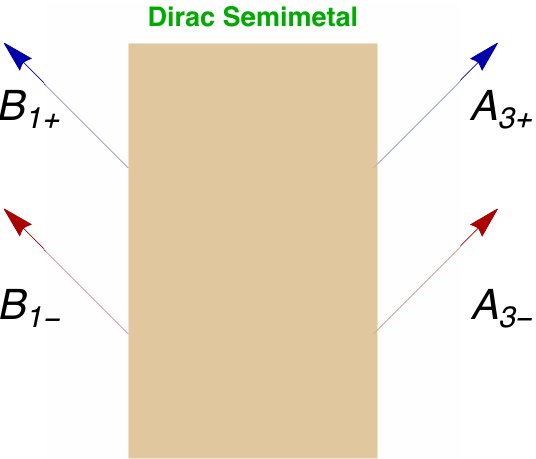}
\caption{The spectral singularity configurations and laser output modes of the Plus-Mode (left panel), 
Minus-Mode (middle panel) and Bimodal case (right panel) in a DSM medium. Waves with amplitudes 
$B_{1+}$ and $A_{3+}$ are output from the left and right sides in the Plus-Mode, amplitudes $B_{1-}$ 
and $A_{3-}$ are for the Minus-Mode, and amplitudes $B_{1+}$, $B_{1-}$, $A_{3+}$, and $A_{3-}$ are for 
the Bimodal case, respectively.}
\label{fig41}
\end{figure}

In this case, there are only outgoing waves of the Plus Mode on the far right and far left sides of the 
system (see Fig.~\ref{fig41}). Here, $B_{1+}$ and $A_{3+}$ are the amplitude of the waves. For this case 
to happen $A_{1+} = A_{1-} = B_{1-} = B_{3+} = A_{3-} = B_{3-} = 0$ must be satisfied, which is possible 
if the condition 
\begin{equation}
\mathbb{M}_{22} = \mathbb{M}_{32} = \mathbb{M}_{42} = 0\,.
\end{equation}
is satisfied. Here $\mathbb{M}_{ij}$ are the components of matrix $\mathbb{M}$. These are the plus-mode 
spectral singularity conditions.

\subsection{Minus-Mode Spectral Singularity Configuration: Minus-Mode Topological DSM Laser}

In this case, the system has only outgoing waves of the Minus-Mode on the far right and far left sides, 
as shown in Fig.~\ref{fig41}. These waves have amplitudes $B_{1-}$ and $A_{3-}$. For this case to occur, 
we must have $A_{1+} = A_{1-} = B_{1+} = B_{3+} = A_{3+} = B_{3-} = 0$. This can only happen if
\begin{equation}
\mathbb{M}_{14} = \mathbb{M}_{24} = \mathbb{M}_{44} = 0 
\end{equation}
is provided for real $k$ values. These are the spectral singularity conditions for the Minus-Mode.

\subsection{Bimodal Spectral Singularity Configuration: Bimodal Topological DSM Laser}

In this case, the system has only outgoing waves of the Bimodal case on the far right and far left 
sides, as shown in Fig.~\ref{fig41}. These waves have amplitudes $B_{1+}$, $B_{1-}$, $A_{3+}$, and 
$A_{3-}$. For this to occur, conditions $A_{1+} = A_{1-} = B_{3+} = B_{3-} = 0$ must be satisfied. This 
is possible if
\begin{equation}
\mathbb{M}_{22} = \mathbb{M}_{24} = \mathbb{M}_{42} = \mathbb{M}_{44} = 0
\end{equation}
is provided for real $k$-values. These are the spectral singularity conditions for the Bimodal 
configuration.

\subsection{Random Spectral Singularity Configurations: Random Topological DSM Lasers}

The setup we have analyzed enables us to create random spectral singularities, making it possible to 
generate random laser beam outputs from either side of the DSM, irrespective of any particular lasing 
mode. This approach illustrates effectiveness of a topological laser configuration for achieving desired 
outcomes, and demonstrates the potential of our method to yield highly diverse results.

As shown in Table~\ref{table3}, 15 different lasing conditions can be achieved in a DSM material. Some 
of these conditions are appropriate for unidirectional lasing, while others are suitable for 
bidirectional lasing or random lasing. Achieving a specific lasing configuration requires meeting 
relevant spectral singularity condition. However, as can be seen in Table~\ref{table3}, it is not 
possible to obtain laser states that only exit from the right side for the wave configurations emitted 
from the left side. Similarly, for the waves emitted from the right side, only laser configurations 
that emerge from the left side are not permitted. In this case, the 15 distinct topological laser states 
observed are actually reduced to 12.

\begin{table}[!ht]
\caption{All potential laser output configurations and conditions from both sides of the DSM slab for an
electromagnetic wave incident from the left side.}
\begin{tabular}{| c | c | c | c |} 
\hline 
Type of Laser & Left Side & Right Side & Spectral Singularity Condition \\ [0.5ex]
\hline \hline & & & \\
Unidirectional Laser from Left (+ Mode) & ~~~~$+$~~~ & ~~~~None~~~ & 
$\mathbb{M}_{12} = \mathbb{M}_{22} = \mathbb{M}_{32} = \mathbb{M}_{42} = 0$ \\
&  &  &\\
\hline & & &\\
Unidirectional Laser from Left ($-$ Mode) & ~~~~$-$~~~ & ~~~~None~~~ & 
$\mathbb{M}_{14} = \mathbb{M}_{24} = \mathbb{M}_{34} = \mathbb{M}_{44} = 0$ \\
& & & \\
\hline & & & $\mathbb{M}_{12} = \mathbb{M}_{22} = \mathbb{M}_{32} = \mathbb{M}_{42} = 0$ \\
Unidirectional Laser from Left (Bimodal) & ~~~~$+$ \& $-$~~~ & ~~~~None~~~ & \\
& & & $\mathbb{M}_{14} = \mathbb{M}_{24} = \mathbb{M}_{34} = \mathbb{M}_{44} = 0$ \\
\hline & & & \\ 
Unidirectional Laser from Right ($+$ Mode) & ~~~~None~~~ & ~~~~$+$~~~ & NOT allowable \\
& & & \\
\hline & & & \\
Unidirectional Laser from Right ($-$ Mode) & ~~~~None~~~ & ~~~~$-$~~~ & NOT allowable \\
& & & \\
\hline & & & \\
Unidirectional Laser from Right (Bimodal)  & ~~~~None~~~ & ~~~~$+$ \& $-$~~~ & NOT allowable \\
& & & \\
\hline & & &\\
Bidirectional Laser ($+$ Mode)             & ~~~~$+$~~~  & ~~~~$+$~~~ & 
$\mathbb{M}_{22} = \mathbb{M}_{32} = \mathbb{M}_{42} = 0$ \\
& & & \\
\hline & & &\\
Bidirectional Laser ($+$ from Left, $-$ from Right) & ~~~~$+$~~~ & ~~~~$-$~~~ & 
$\mathbb{M}_{12} = \mathbb{M}_{22} = \mathbb{M}_{42} = 0$ \\
& & & \\
\hline & & &\\
Bidirectional Laser~~($-$~from Left, $+$ from Right) & ~~~~$-$~~~ & ~~~~$+$~~~ & 
$\mathbb{M}_{24} = \mathbb{M}_{34} = \mathbb{M}_{44} = 0$\\
& & & \\
\hline & & & \\
Bidirectional Laser ($-$ Mode) & ~~~~$-$~~~ & ~~~~$-$~~~ & 
$\mathbb{M}_{14} = \mathbb{M}_{24} = \mathbb{M}_{44} = 0$ \\
& & & \\
\hline & & & \\
Bidirectional Laser ($+$ from Left, $+$ \& $-$ from Right) & ~~~~$+$~~~ & ~~~~$+$ \& $-$~~~ & 
$\mathbb{M}_{22} = \mathbb{M}_{42} = 0$ \\
& & & \\
\hline & & & \\
Bidirectional Laser ($-$ from Left, $+$ \& $-$ from Right) & ~~~~$-$~~~ & ~~~~$+$ \& $-$~~~ & 
$\mathbb{M}_{24} = \mathbb{M}_{44} = 0$ \\
& & & \\
\hline & & & $\mathbb{M}_{22} = \mathbb{M}_{24} = 0$ \\
Bidirectional Laser ($+$ \& $-$ from Left, $+$ from Right) & ~~~~$+$ \& $-$~~~ & ~~~~$+$~~~ & 
$\mathbb{M}_{32} = \mathbb{M}_{34} = 0$ \\
& & & $\mathbb{M}_{42} = \mathbb{M}_{44} = 0$ \\
\hline & & & $\mathbb{M}_{12} = \mathbb{M}_{14} = 0$ \\
Bidirectional Laser ($+$ \&$-$ from Left, $-$ from Right) & ~~~~$+$ \& $-$~~~ & ~~~~$-$~~~ & 
$\mathbb{M}_{22} = \mathbb{M}_{24} = 0$ \\
& & & $\mathbb{M}_{42} = \mathbb{M}_{44} = 0$ \\
\hline & & & \\
Bidirectional Laser (Bimodal) & ~~~~$+$ \& $-$~~~ & ~~~~$+$ \& $-$~~~ & 
$\mathbb{M}_{22} = \mathbb{M}_{24} = \mathbb{M}_{42} = \mathbb{M}_{44} = 0$ \\
& & &\\
\hline 
\end{tabular}
\label{table3}
\end{table}

\begin{figure}[!hbt]
\includegraphics[width=8cm]{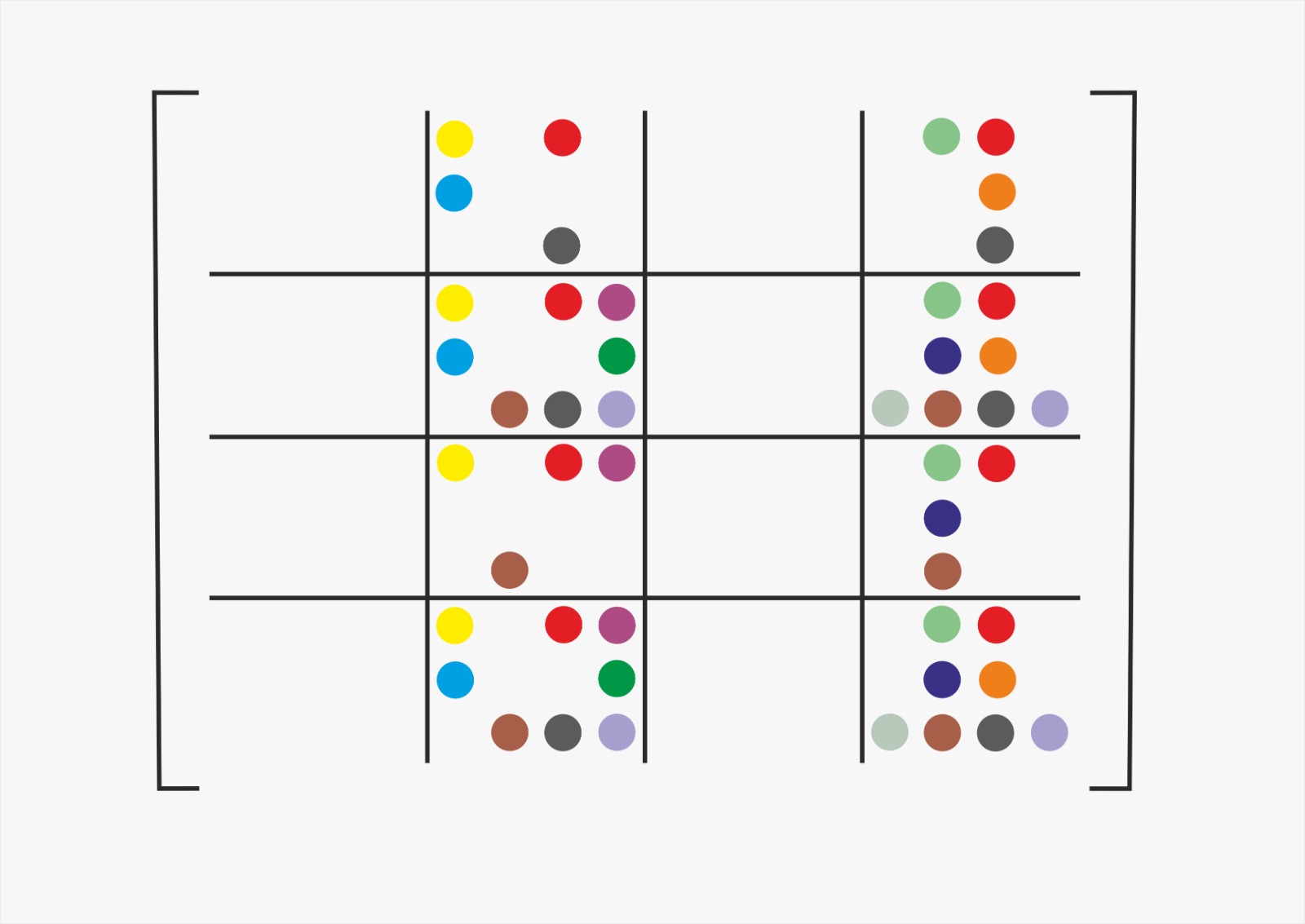}
\caption{A diagram of topological DSM laser types within the components of the transfer matrix for a DSM environment. 
It illustrates the laser generation conditions for each topological DSM laser type listed in Table~\ref{table3}, 
with different colors representing different laser types. Notably, only the second and fourth columns of 
the transfer matrix generate laser beams.}
\label{transfermatr}
\end{figure}

Figure~(\ref{transfermatr}) shows components of the transfer matrix. It is manifestly seen which 
components of the transfer matrix provide the conditions given in Table~\ref{table3}, corresponding to 
distinct laser configurations. Figure~(\ref{transfermatr}) reveals that the configurations presented in 
Table~\ref{table3} originate from diverse combinations of components in the second and last columns of 
the transfer matrix. This highlights that these columns play a pivotal role in determining all distinct 
laser configurations in a DSM slab.

\begin{figure*}[!hbt]
\begin{tikzpicture} 
\node[anchor=north west,inner sep=0pt] at (0,0){\includegraphics[width=5cm]{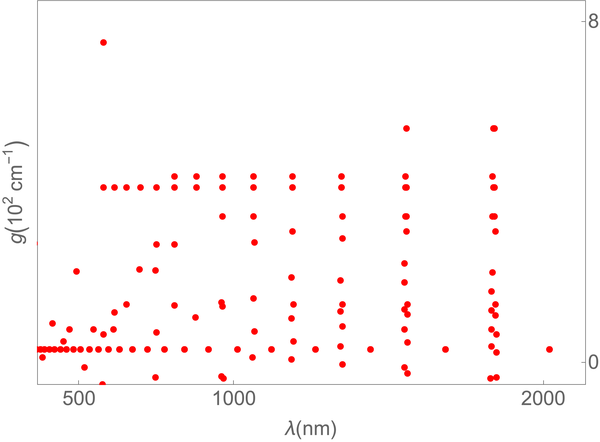}};
\node[font=\sffamily\bfseries\large] at (19ex,-2ex) {(a)}; \draw[orange, thick,->] (4.9,-3.3) -- (5.8,-4);
\end{tikzpicture} 
\begin{tikzpicture}  
\node[anchor=north west,inner sep=0pt] at (0,0){\includegraphics[width=5cm]{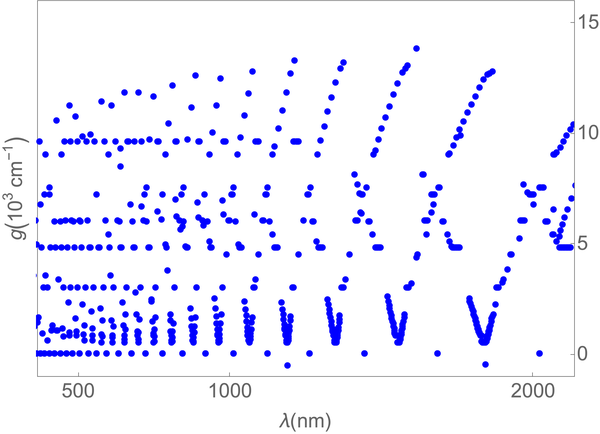}};
\node[font=\sffamily\bfseries\large] at (19ex,-2ex) {(b)}; \draw[orange, thick,->] (2.6,-3.6) -- (2.6,-4);
\end{tikzpicture}
\begin{tikzpicture} 
\node[anchor=north west,inner sep=0pt] at (0,0){\includegraphics[width=5cm]{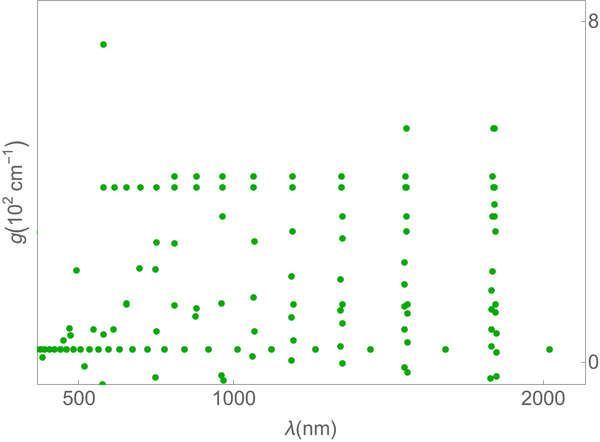}}; 
\node[font=\sffamily\bfseries\large] at (19ex,-2ex) {(c)}; \draw[orange, thick,->] (0.3,-3.3) -- (-0.7,-4);
\end{tikzpicture}\ 
\begin{tikzpicture} 
\node[anchor=north west,inner sep=0pt] at (0,0){\includegraphics[width=5cm]{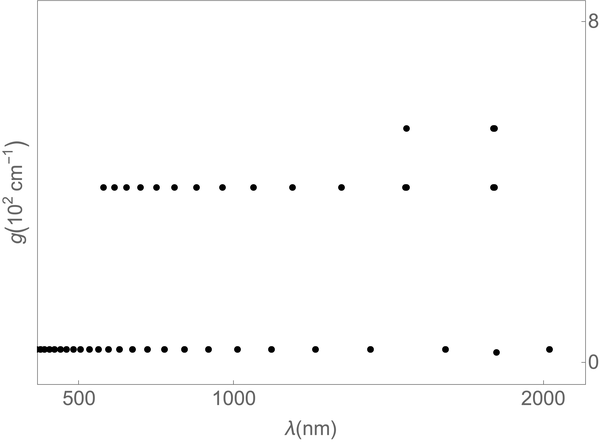}};
\node[font=\sffamily\bfseries\large] at (19ex,-4ex) {(d)};
\end{tikzpicture}
\caption{The spectral singularities are displayed over the $\lambda - g$ parameters of the Plus Mode 
configuration. Panels (a), (b) and (c) show the real zero values of components of the transfer matrix 
separately, while intersection points of these points are shown in panel (d). Graphs are plotted using 
the data obtained from Eq.~(\ref{specifications}).} 
\label{figg1}
\end{figure*}
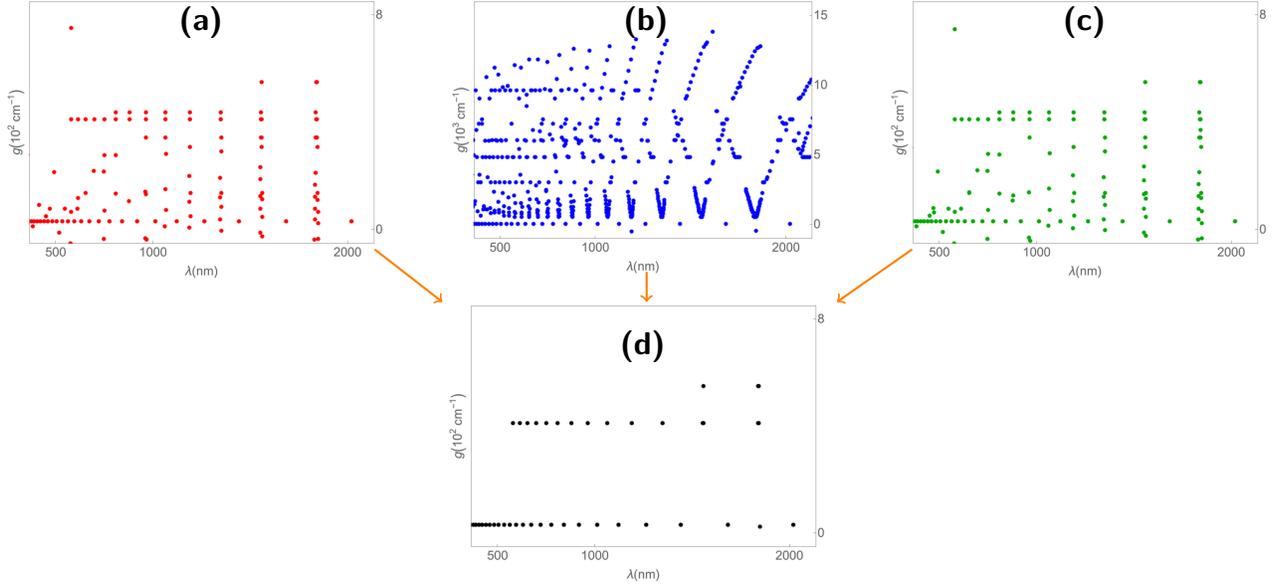

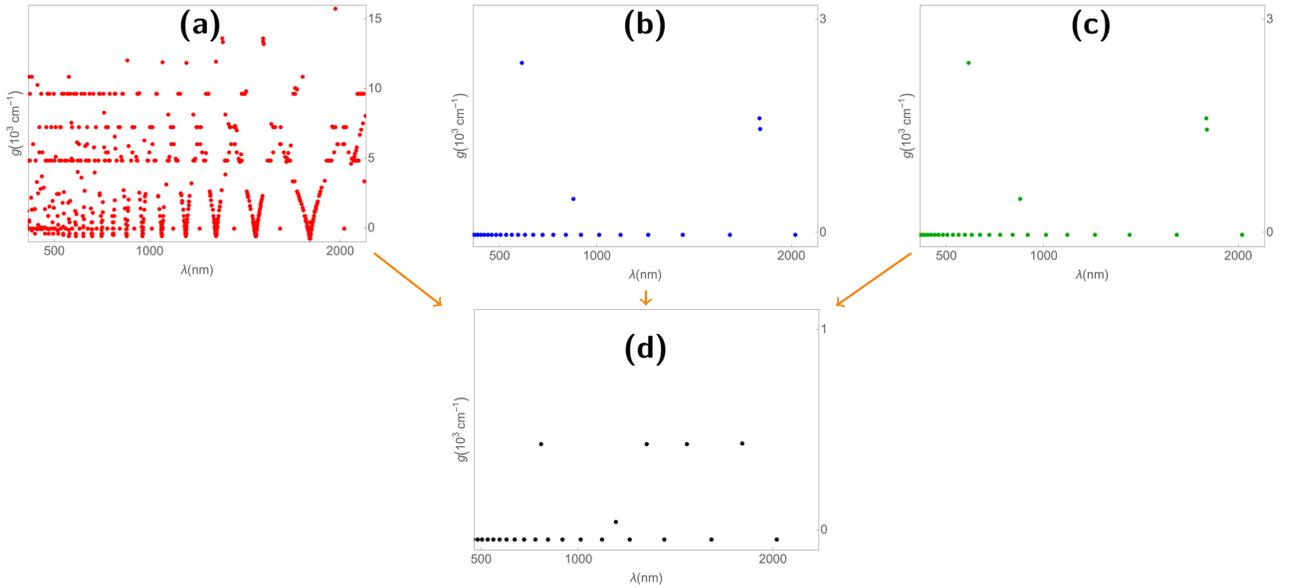
\begin{figure}[!hbt]
\begin{tikzpicture} 
\node[anchor=north west,inner sep=0pt] at (0,0){\includegraphics[width=5cm]{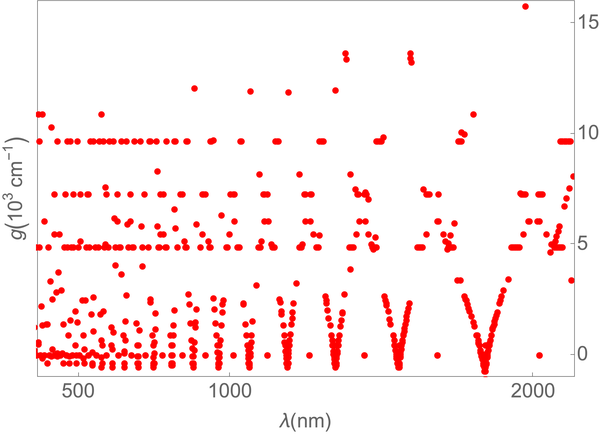}};
\node[font=\sffamily\bfseries\large] at (19ex,-2ex) {(a)}; \draw[orange, thick,->] (4.9,-3.3) -- (5.8,-4);
\end{tikzpicture} 
\begin{tikzpicture}  
\node[anchor=north west,inner sep=0pt] at (0,0){\includegraphics[width=5cm]{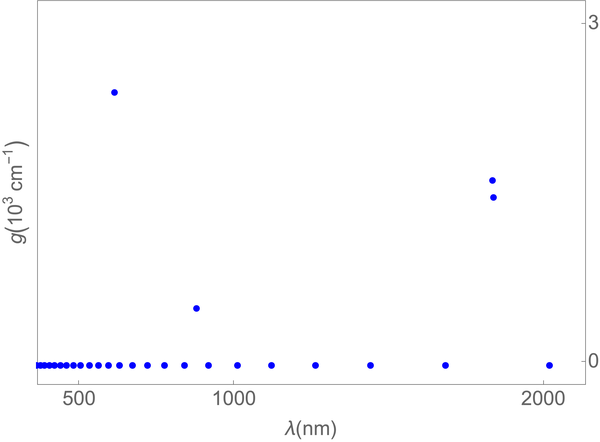}};
\node[font=\sffamily\bfseries\large] at (19ex,-2ex) {(b)}; \draw[orange, thick,->] (2.6,-3.8) -- (2.6,-4);
\end{tikzpicture}
\begin{tikzpicture} 
\node[anchor=north west,inner sep=0pt] at (0,0){\includegraphics[width=5cm]{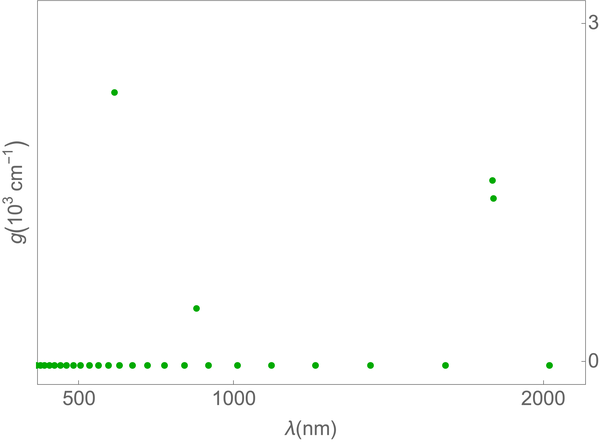}}; 
\node[font=\sffamily\bfseries\large] at (19ex,-2ex) {(c)}; \draw[orange, thick,->] (0.2,-3.3) -- (-0.8,-4);
\end{tikzpicture}\ 
\begin{tikzpicture} 
\node[anchor=north west,inner sep=0pt] at (0,0){\includegraphics[width=5cm]{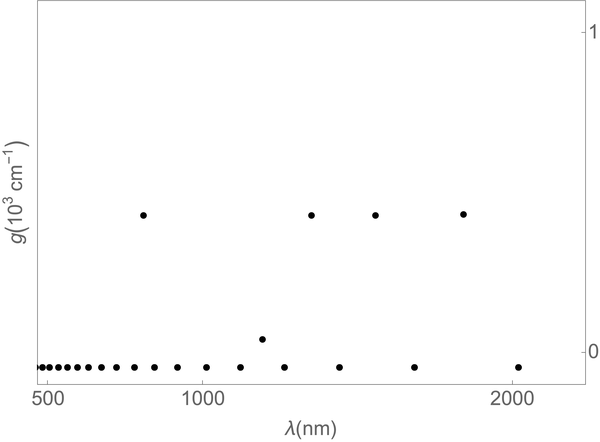}};
\node[font=\sffamily\bfseries\large] at (19ex,-4ex) {(d)};
\end{tikzpicture}
\caption{Spectral singularities as a function of the $\lambda - g$ parameters in the Minus Mode 
configuration. Panels (a), (b), and (c) show the real zero values of individual components of the 
transfer matrix, while intersection points of these zeros are presented in panel (d) below. These graphs 
are generated using the data provided in Eq.~(\ref{specifications}).}
\label{figg2}
\end{figure}

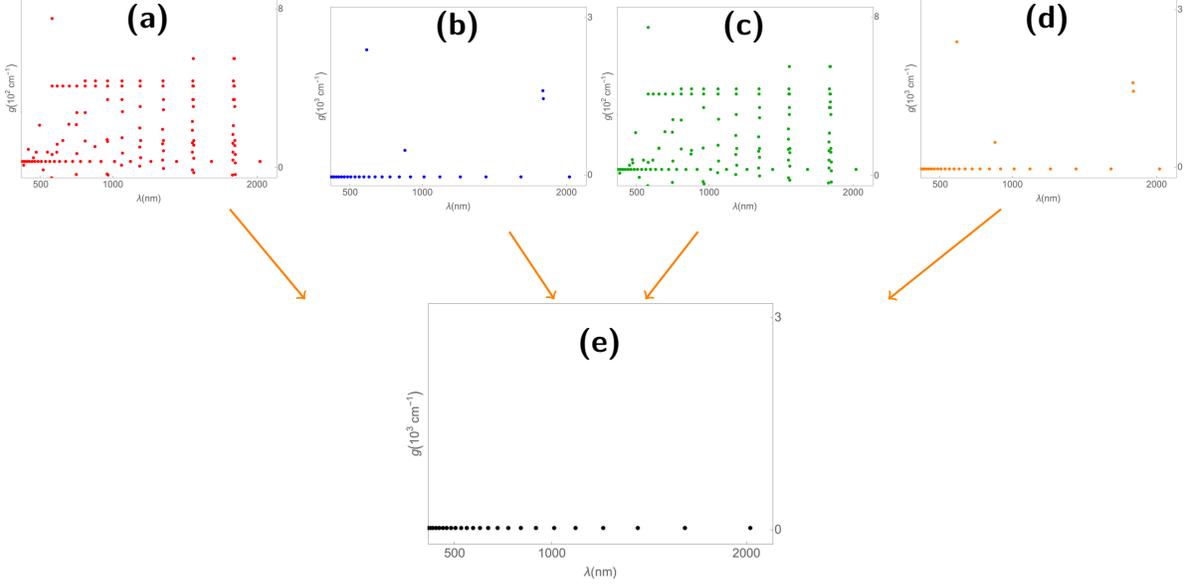
\begin{figure}[!hbt]
\begin{tikzpicture} 
\node[anchor=north west,inner sep=0pt] at (0,0){\includegraphics[width=3.7cm]{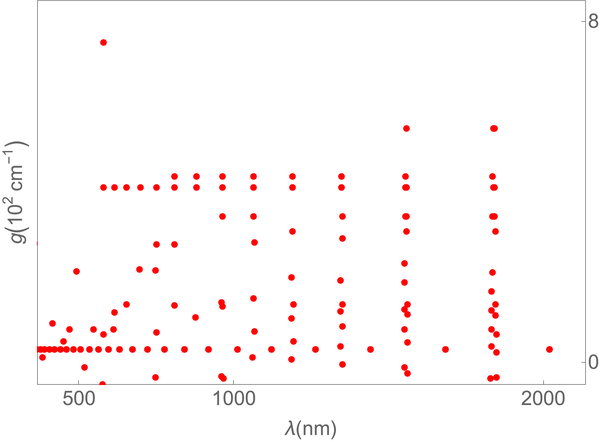}};
\node[font=\sffamily\bfseries\large] at (14ex,-2ex) {(a)}; \draw[orange, thick,->] (3,-2.8) -- (4,-4);
\end{tikzpicture} 
\begin{tikzpicture}  
\node[anchor=north west,inner sep=0pt] at (0,0){\includegraphics[width=3.7cm]{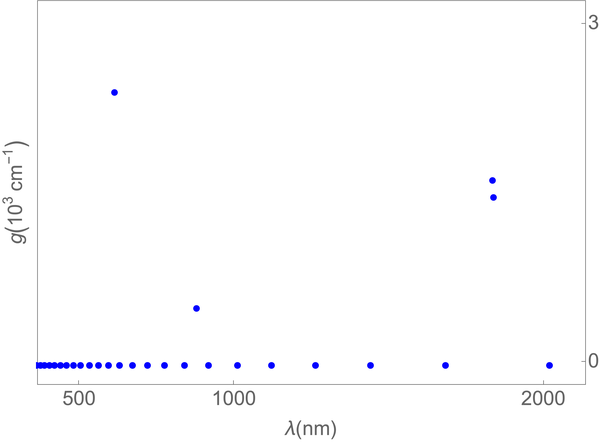}};
\node[font=\sffamily\bfseries\large] at (14ex,-2ex) {(b)}; \draw[orange, thick,->] (2.6,-3) -- (3.2,-3.9);
\end{tikzpicture}
\begin{tikzpicture} 
\node[anchor=north west,inner sep=0pt] at (0,0){\includegraphics[width=3.7cm]{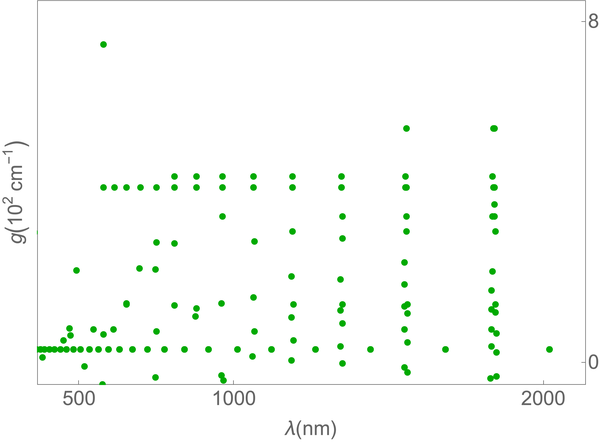}}; 
\node[font=\sffamily\bfseries\large] at (14ex,-2ex) {(c)}; \draw[orange, thick,->] (1.3,-3) -- (0.6,-3.9);
\end{tikzpicture}
\begin{tikzpicture} 
\node[anchor=north west,inner sep=0pt] at (0,0){\includegraphics[width=3.7cm]{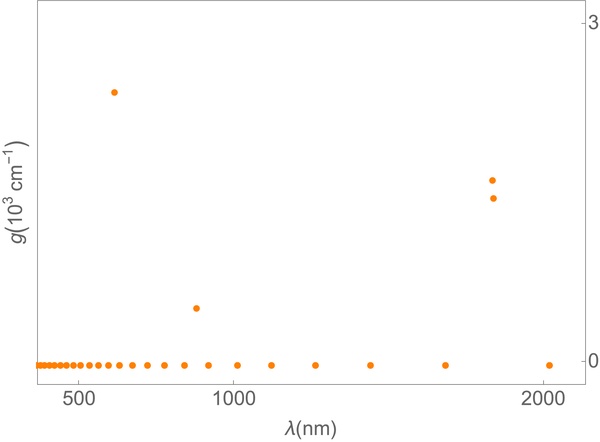}}; 
\node[font=\sffamily\bfseries\large] at (14ex,-2ex) {(d)}; \draw[orange, thick,->] (1.3,-2.8) -- (-0.2,-4);
\end{tikzpicture}\\
\begin{tikzpicture} 
\node[anchor=north west,inner sep=0pt] at (0,0){\includegraphics[width=5cm]{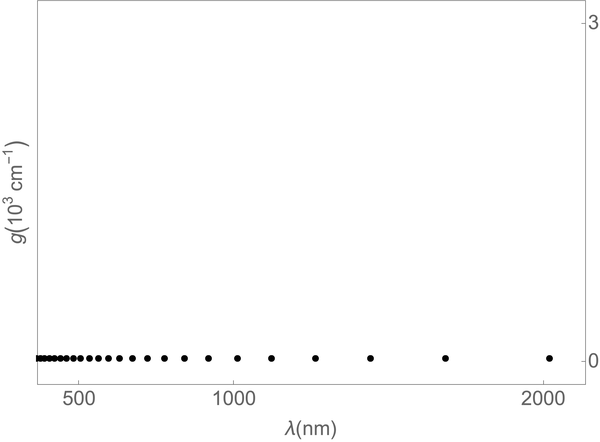}};
\node[font=\sffamily\bfseries\large] at (19ex,-4ex) {(e)};
\end{tikzpicture}
\caption{The spectral singularities as a function of the $\lambda - g$ parameters for the bimodal 
configuration. Panels (a), (b), (c), and (d) display the real zero values of individual components of 
the transfer matrix, while the intersection points of these zeros are shown in panel (e). These graphs 
are based on the data provided by Eq.~(\ref{specifications}).} 
\label{figg3}
\end{figure}

To understand how a topological DSM laser is shaped based on system parameters, we first need to determine the laser 
type and apply relevant conditions. Among the 12 laser types, we will focus on only three key ones: 
Plus-Mode, Minus-Mode, and Bimodal topological DSM lasers.
0
The DSM slab system is influenced by several parameters, such as the gain coefficient $g$, wavelength 
$\lambda$ of the incoming wave, incident angle $\phi$, slab thickness $L$, and material type, 
characterized by $\eta$. Desired conditions are determined by optimal interdependence of these 
parameters. For convenience, we select Na$_3$Bi as the DSM material, with its properties given in 
Eq.~(\ref{specifications}), relationship between $g$ and $\lambda$ for three different laser types is 
shown in Figs.~(\ref{figg1}), (\ref{figg2}), and (\ref{figg3}).

Figure~\ref{figg1} presents conditions for the Plus Mode laser type, where only Plus Mode waves can exit 
from both the right and left sides of the DSM slab. The colored points in panels (a), (b), and (c) 
represent the spectral singularity points, corresponding to real zeros of $\mathbb{M}_{22}$, 
$\mathbb{M}_{32}$ and $\mathbb{M}_{42}$, respectively. Black spectral singularity points in panel (d) 
represent the common values where these zeros occur simultaneously. In other words, the system will lase 
in Plus-Mode from both sides of the DSM slab at wavelengths indicated by the black points. As shown in 
panel (d), multiple wavelengths can correspond to the same gain value, or a single wavelength can 
correspond to multiple gain values. This is a significant result highlighting topological nature of the 
topological DSM laser. Despite the material having a fixed $\theta$ value, the behavior of laser remains robust to 
changes in the system parameters. This demonstrates that the DSM laser is indeed a topological laser.

We observe similar behavior in the Minus-Mode DSM laser shown in Fig.~(\ref{figg2}). In this case, laser 
beams are emitted from both sides of the DSM laser. Different colors in the upper part of the figure 
represent the real zero points of the $\mathbb{M}_{14}$, $\mathbb{M}_{24}$ and $\mathbb{M}_{44}$ 
components of the transfer matrix, respectively. Intersections of these points are marked in black in 
the lower panel. Additionally, a robust topological laser can be achieved for a wide range of parameter 
values in the Minus Mode. In this case, the wavelength stability is lower compared to the Plus Mode. 
However, it is evident that the system can maintain the same gain value across different wavelengths.

The final type of DSM laser we will examine is the bimodal configuration, which outputs both Plus and 
Minus Mode waves from both sides, as shown in Fig.~(\ref{figg3}). The colored points in the upper panels 
represent the spectral singularity points corresponding to the real zeros of the $\mathbb{M}_{22}$, 
$\mathbb{M}_{24}$, $\mathbb{M}_{42}$  and $\mathbb{M}_{44}$ components of the transfer matrix. To achieve 
laser output in both modes from both sides, common intersection points of these spectral singularities 
must be employed. These intersection points are indicated in black in the lower panel (e). As shown, the 
system maintains robustness of the gain value, but the gain decreases at the same wavelength. However, 
the gain remains robust across different wavelengths. We can thus conclude that robustness of the 
Plus-Mode laser is stronger than that of the Minus and Bimodal configurations. Nevertheless, topological 
characteristics can be observed in all three cases.

Similar analyses can be conducted for the remaining nine cases to explore interconnected topological 
characteristics of $g$ and $\lambda$. These analyses may unveil interesting results that have not yet 
been reported. However, to avoid going off-topic, we will now focus on the effect of another important 
parameter, incident angle $\phi$ on the gain value.

\begin{figure}[!hbt]
\begin{tikzpicture} 
\node[anchor=north west,inner sep=0pt] at (0,0){\includegraphics[width=5.7cm]{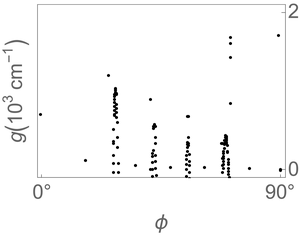}};
\node[font=\sffamily\bfseries\large] at (8ex,-2.5ex) {(a)}; 
\end{tikzpicture} 
\begin{tikzpicture}
\node[anchor=north west,inner sep=0pt] at (0,0){\includegraphics[width=5.7cm]{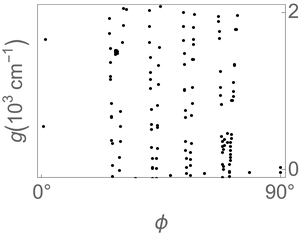}};
\node[font=\sffamily\bfseries\large] at (8ex,-2.5ex) {(b)};
\end{tikzpicture}
\begin{tikzpicture} 
\node[anchor=north west,inner sep=0pt] at (0,0){\includegraphics[width=5.7cm]{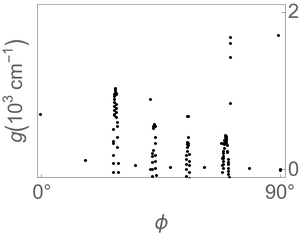}}; 
\node[font=\sffamily\bfseries\large] at (8ex,-2.5ex) {(c)}; 
\end{tikzpicture}
\caption{The spectral singularity points of the Plus Mode [panel (a)], Minus Mode [panel (b)] and 
Bimodal [panel (c)] cases in the $\lambda - g$ plane. These points are found by the intersections of 
points created by the conditions that generate the relevant laser types, as we have shown in 
Figs.~(\ref{figg1}), (\ref{figg2}) and (\ref{figg3}). These graphs are based on the data provided by 
Eq.~(\ref{specifications}).} 
\label{figg4}
\end{figure}
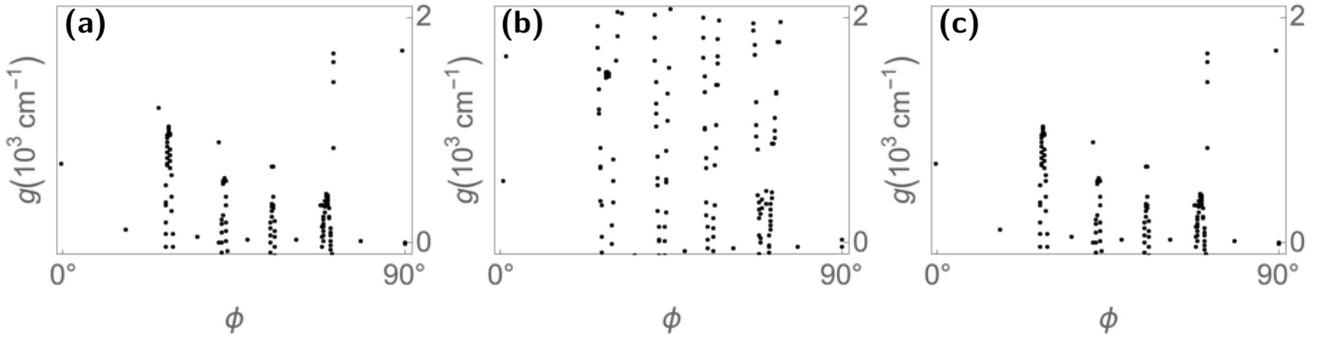

Figure~(\ref{figg4}) shows how the gain coefficient $g$ changes with the incident angle $\phi$ for 
Plus-Mode, Minus-Mode and Bimodal cases. As can be clearly seen in the figures, it is observed that the 
gain value and incident angle parameters change in accordance with the topological character of the DSM 
environment. This is not surprising since this material is topological. The results we found confirm this 
finding.

It is possible to make similar analysis for other parameters, but we will not discuss them in order to 
avoid further distraction and because the most important parameters for our system are the gain 
coefficient, wavelength and incidence angle. All these different configurations occur because the 
electromagnetic interaction in the DSM medium is dichroic in nature, which is due to the constant axion 
term that such a medium contains.

\section{Induced Surface Current $\vec{J}_{\theta}$ and its Behavior at Spectral Singularities}

Before concluding our discussion, one more important issue needs to be addressed: the axion-induced 
surface current(s), denoted as $\vec{J}_{\theta}$. Main reason for the formation of this surface current 
is that, unlike in Weyl semimetals, the $\theta$-parameter in DSM undergoes a discontinuous jump at the 
surface. As demonstrated in Appendix B, the axionic surface current is given by 
$\vec{J}_{\theta} = -\beta \mu \vec{\nabla} \theta \times \vec{E}$. Using the expression for 
$\vec{\nabla} \theta = \pi [\delta(z) \Theta(L-z) + \Theta(z) \delta(L-z)]\, \hat{e}_k$, which corresponds 
to a DSM slab, and the electric field provided in Table~\ref{t1}, we can derive the following expression 
for the axion-induced surface current:
\begin{equation}
\vec{J}_{\theta} = \frac{(1-i) \beta \mu \pi}{4} [\delta(z) \Theta(L-z) + \Theta(z) \delta(L-z)] 
\left[\mathcal{F}_{+}^{+}(z) - i \, \mathcal{F}_{-}^{+}(z)\right] e^{ik_x x} \hat{e}_x.
\end{equation}
As is evident from this expression, the axion-induced current(s) arise at the surfaces $z = 0$ and 
$z = L$, and they flow in the $x$-direction. To better understand general characteristics of these 
currents, we will examine three different configurations: the Plus-Mode, Minus-Mode, and Bimodal Cases.

By revisiting the spectral singularity condition for the Plus-Mode, we obtain 
$A_{3+} = \mathbb{M}_{12} B_{1+}$. Thus, it becomes clear that the surface current takes the following 
simplified form:
\begin{align}
	J_{\theta}(x, z) : = \frac{(1-i)\beta \mu \pi e^{ik_x x}\,B_{1+}}{4}
	\begin{cases} 
		1, \hskip 2cm z = 0,\\
		\mathbb{M}_{12}\, e^{ik_z L}, \qquad z = L.
	\end{cases} \label{Fpm2}
\end{align}
Similarly, using the spectral singularity condition for the Minus-Mode, one gets the result 
$A_{3-} = \mathbb{M}_{34} B_{1-}$. Hence, it becomes evident that the surface current simplifies to the 
following form:
\begin{align}
	J_{\theta}(x, z) : = -\frac{(1+i)\beta \mu \pi e^{ik_x x}\,B_{1-}}{4}
	\begin{cases} 
		1, \hskip 2cm z = 0,\\
		\mathbb{M}_{34}\, e^{ik_z L}, \qquad z = L.
	\end{cases} \label{Fpm3}
\end{align}
Finally, by applying the spectral singularity condition for the Bimodal case, we find the expressions 
$A_{3+} = \mathbb{M}_{12} B_{1+} + \mathbb{M}_{14} B_{1-}$ and $A_{3-} = \mathbb{M}_{32} B_{1+} + 
\mathbb{M}_{34} B_{1-}$. Therefore, bimodal surface currents at $z = 0$ and $z = L$ turn out to be 
\begin{align}
J_{\theta}(x, z) : = \frac{(1-i)\beta \mu \pi e^{ik_x x}}{4}
\begin{cases} 
[B_{1+}-iB_{1-}], \hskip 5.5cm z = 0,\\
[B_{1+}(\mathbb{M}_{12}-i\mathbb{M}_{32}) + B_{1-}(\mathbb{M}_{14}-i\mathbb{M}_{34})]\, e^{ik_z L}, \qquad z = L.
\end{cases} \label{Fpm4}
\end{align}

\begin{figure}[!hbt]
  \begin{tikzpicture} 
  \node[anchor=north west,inner sep=0pt] at (0,0){\includegraphics[width=5.7cm]{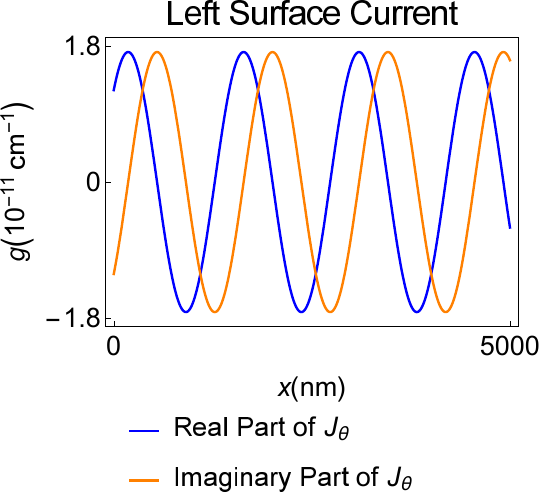}};
  \node[font=\sffamily\bfseries\large] at (10ex,-1ex) {(a)}; 
  \end{tikzpicture} 
  \begin{tikzpicture}  
  \node[anchor=north west,inner sep=0pt] at (0,0){\includegraphics[width=5.7cm]{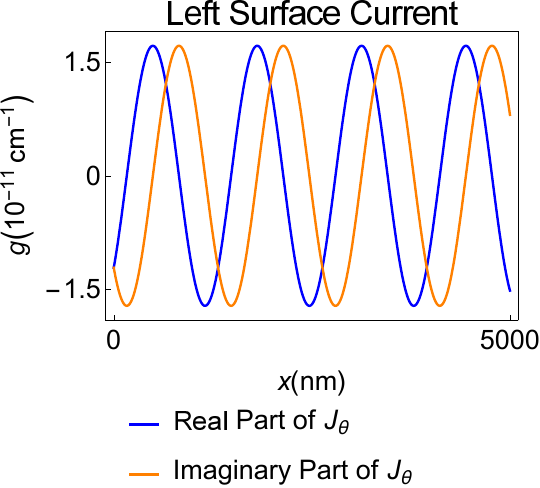}};
  \node[font=\sffamily\bfseries\large] at (10ex, -1ex) {(c)};
  \end{tikzpicture}
  \begin{tikzpicture} 
  \node[anchor=north west,inner sep=0pt] at (0,0){\includegraphics[width=5.7cm]{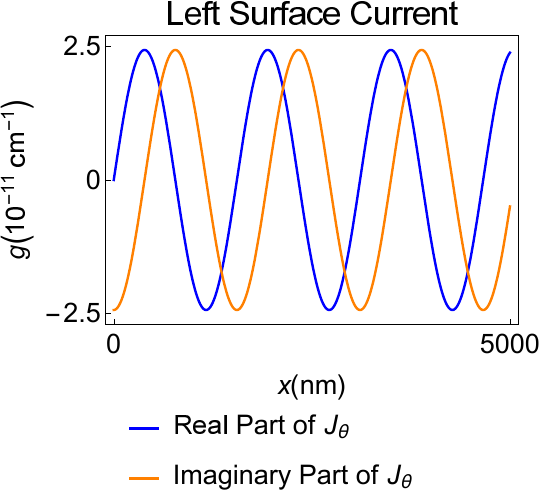}}; 
  \node[font=\sffamily\bfseries\large] at (10ex,-1ex) {(e)}; 
  \end{tikzpicture}\\
  \begin{tikzpicture} 
  \node[anchor=north west,inner sep=0pt] at (0,0){\includegraphics[width=5.7cm]{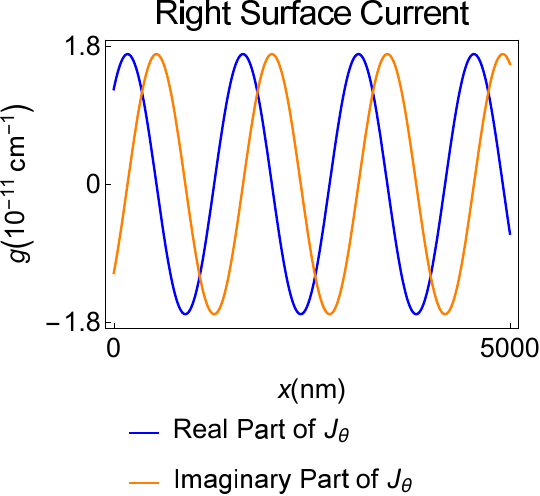}};
  \node[font=\sffamily\bfseries\large] at (10ex,-1ex) {(b)}; 
  \end{tikzpicture} 
  \begin{tikzpicture}  
  \node[anchor=north west,inner sep=0pt] at (0,0){\includegraphics[width=5.7cm]{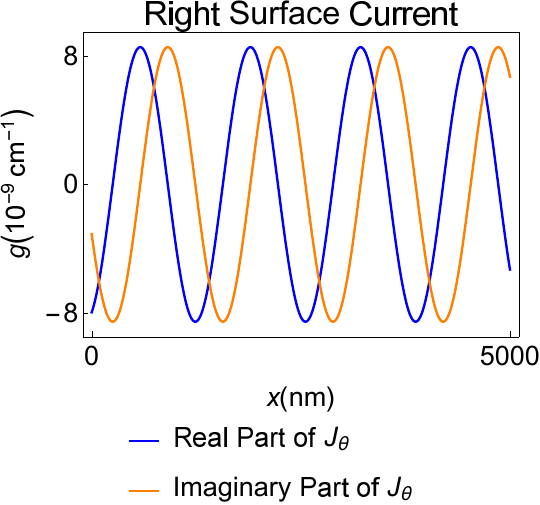}};
  \node[font=\sffamily\bfseries\large] at (8ex,-1ex) {(d)};
  \end{tikzpicture}
  \begin{tikzpicture} 
  \node[anchor=north west,inner sep=0pt] at (0,0){\includegraphics[width=5.7cm]{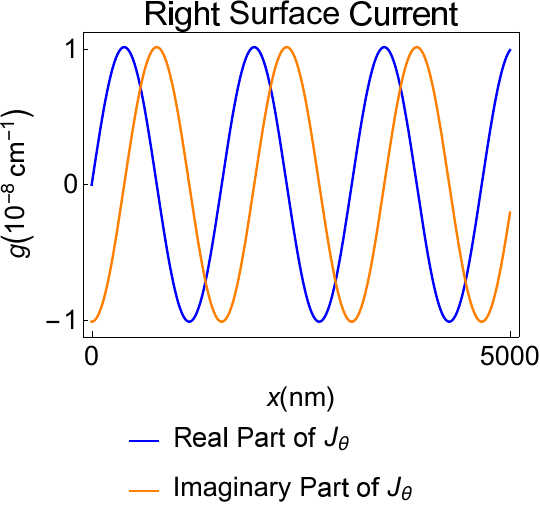}}; 
  \node[font=\sffamily\bfseries\large] at (8ex,-1ex) {(f)}; 
  \end{tikzpicture}
 \caption{Current configurations on the left and right surfaces of the DSM slab in three different modes: 
Plus-Mode, Minus-Mode, and Bimodal Case. Panels (a) and (b) show surface currents in Plus-Mode, panels 
(c) and (d) depict surface currents in the Minus-Mode, and panels (e) and (f) represent surface currents 
in the Bimodal Case.} \label{figgc}
\end{figure}
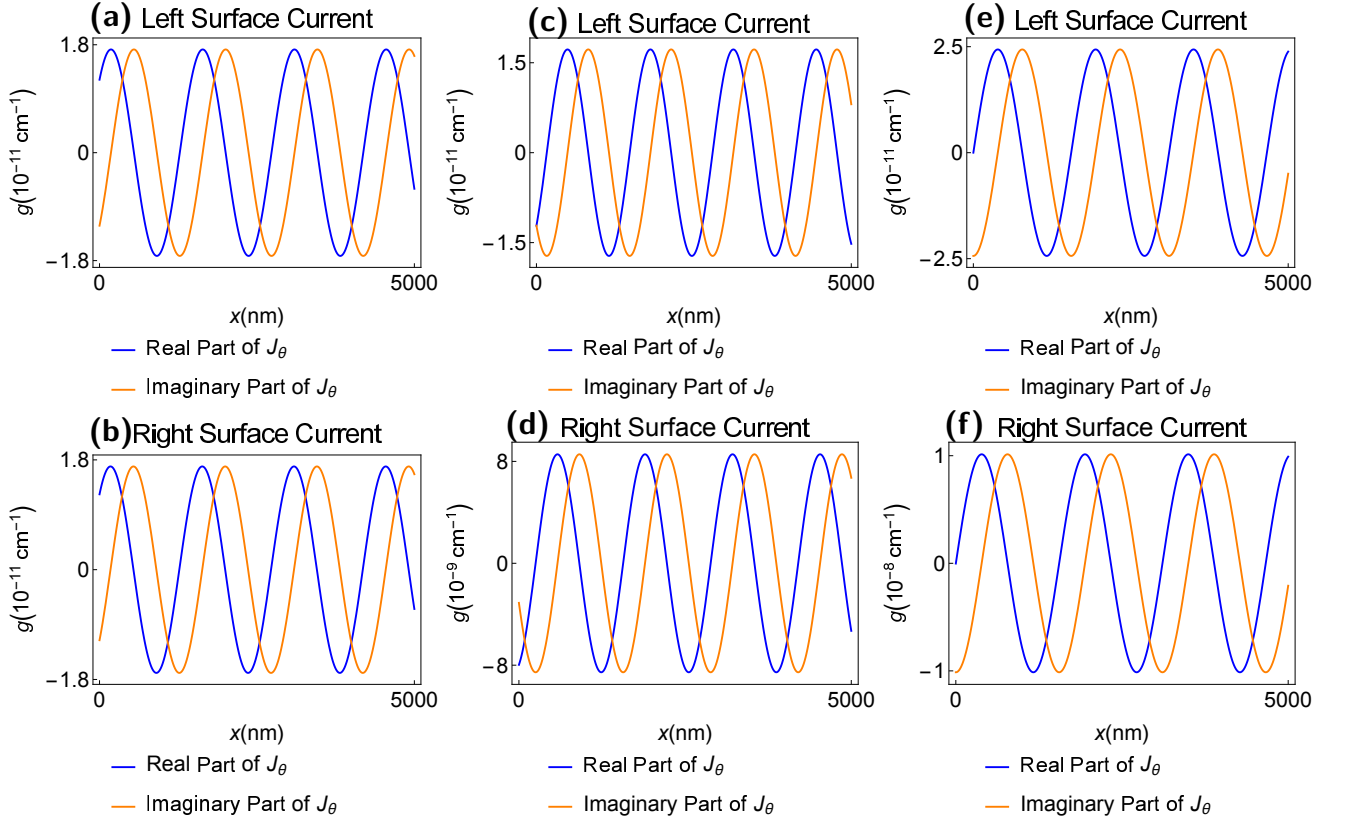

As shown in Fig.~(\ref{figgc}), surface current configurations induced on the left and right surfaces of 
the DSM plate are consistent with the expressions derived above. These currents are generated at the 
spectral singularity points (SSP) within the DSM. These spectral singularity points are 
$\textrm{SSP1} \in \{g= 47.166~\textrm{cm}^{-1}, \lambda = 1261.75~\textrm{nm}, \phi = 60^{\circ}, 
L = 5~\mu \textrm{m}\}$, $\textrm{SSP2} \in \{g= 4831.520~\textrm{cm}^{-1}, \lambda = 1140.33~\textrm{nm}, 
\phi= 60^{\circ}, L= 5~\mu$~m and $\textrm{SSP3} \in \{g= 32.678~\textrm{cm}^{-1}, 
\lambda = 1345.87~\textrm{nm}, \phi= 60^{\circ}, L= 5~\mu \textrm{m}\}$, respectively. It should be noted 
that, apart from the spectral singularity points, a clear phase difference occurs between the left and 
right surface currents at a different point. However, this phase difference vanishes at the spectral 
singularity points, as demonstrated in Fig.~(\ref{figgc}). In addition, current on the right surface is 
greater than that on the left surface in all three modes.

We know that due to its semimetallic nature, a current $\vec{J}_f$ flows within a DSM material. However, 
we will not discuss the details of this current here.

\section{Concluding Remarks}

The unique properties of semimetallic materials make them particularly fascinating, and their 
electromagnetic interactions and optical applications have sparked considerable interest. In our study, 
we aimed to shed light on these intriguing aspects and achieved some very interesting findings. 
Fundamental characteristics of Dirac and Weyl semimetals arise from the presence of the theta term in 
their structure. Among these, Dirac semimetals (DSMs) stand out, as the origin of the theta term lies in 
the single Dirac cone and a fixed value of the theta parameter. By exploring this distinctive feature 
from a non-Hermitian perspective, our research led to surprising results that have not been explicitly reported in previous studies. One of the most notable discoveries is that these materials inherently exhibit 
dichroism, leading to a multidimensional nature in their interactions with electromagnetic waves.

Through our examination of the TE mode solutions, we observed that the dichroism effect exhibited by DSM 
materials is consistent with experimental findings in the literature, where it has been noted in certain 
DSM candidate materials. The dichroism effect, as we have shown, also leads to birefringence in these 
materials. As a consequence, an electromagnetic wave incident on the material for scattering becomes 
effectively 2D within the material. This alters the transfer matrix, making it $4 \times 4$ in dimension, 
which is counter to what one might initially expect. Interestingly, this outcome mirrors a phenomenon we 
encountered in Weyl semimetals, though it arises from a different cause-Faraday rotation. Despite the 
similarities, it is important to emphasize that the underlying mechanisms for these effects are 
fundamentally distinct.

By connecting the scattering solutions of DSMs to non-Hermitian physics and associating them with spectral 
singularities, we open the door to the possibility of creating a topological laser from these materials. 
In this study, we explored this potential by investigating how DSMs could serve as the foundation for a 
topological laser. As is well known, achieving a laser effect requires the presence of spectral 
singularities, which can be obtained through the transfer matrix. These singularities correspond to the 
laser threshold conditions that produce zero-width resonance states. Notably, we discovered that DSMs can 
generate laser output in twelve distinct ways. This phenomenon arises from two different modes present in 
DSMs-namely, the Plus and Minus Modes-which can each produce different results in a preferred manner. 
This result, which has not been observed in literature, provides new insight into the behavior of these 
materials. By examining these different laser modes in detail, we believe we can uncover previously 
unknown characteristics of DSM materials and open up new possibilities for diverse laser applications.

These materials also give rise to axion-induced surface currents, a phenomenon that holds significant 
importance for a variety of potential applications. Through our analysis of these currents, we 
demonstrated that they appear in phase and in the same direction on both surfaces of the DSM slab. The 
results of our study, along with the promising application areas, contribute to a deeper understanding 
of the unique properties of these materials. This insight is likely to generate increased interest and 
pave the way for a wide range of innovative applications in the near future.

\appendix

\section{Appendix A. Derivation of Maxwell Equations and Modified Maxwell Equations in Axion 
Electrodynamics} \label{S2}

In a DSM at low energy limits, a spatially varying axion term plays an important role in determining 
electromagnetic fields. Total action of the corresponding DSM plate system is defined as the sum of the 
conventional and axionic terms as follows: $S = S_0 + S_{\theta}$.
\begin{align}
S_0 &= \int \left\{-\frac{1}{4\mu_0}F_{\mu\nu}F^{\mu\nu} + \frac{1}{2}F_{\mu\nu} \mathcal{P}^{\mu\nu} - 
J^{\mu}A_{\mu}\right\} d^3x\,dt,\\
S_{\theta} &= \frac{\alpha}{8\pi\mu_0}\int \left\{\theta(\vec{r}, t)\varepsilon^{\mu\nu\alpha\beta}
F_{\mu\nu}F_{\alpha\beta}\right\} d^3x\,dt,
\end{align}
Here, $\mathcal{P}^{\mu\nu}$ represents the electric polarization and magnetization given by 
$\mathcal{P}^{0i} = cP^{i}$ and $\mathcal{P}^{ij} = -\varepsilon^{ijk} M_{k}$, respectively. $A_{\mu}$ 
is the 4-vector potential, and $F_{\mu\nu}$ is the completely antisymmetric electromagnetic field 
strength tensor. The axion term, which depends on space and time, is given by 
$\theta(\vec{r}, t) = \pi \,\Theta(z)\,\Theta(L-z)$, where $\Theta(z)$ is the Heaviside step function. 
For a DSM, the $\theta$ term is time-independent and takes a constant value of $\pi$ within a limited 
region. Taking variation of the action with respect to $A_\mu$, following equations of motion are 
obtained.
\begin{equation}
 - \frac{1}{\mu_0}\partial_{\nu}F^{\mu\nu} + \partial_{\nu}\mathcal{P}^{\mu\nu} + 
\frac{\alpha}{2\pi\mu_0}\varepsilon^{\mu\nu\alpha\beta}\partial_{\nu}\left(\Theta 
F_{\alpha\beta}\right) = J^{\mu}
 \end{equation}
Writing this equation yields modified Maxwell equations given in Eqs.~(\ref{1}), (\ref{2}), (\ref{eq3}),  
and (\ref{eq4}) in the presence of the axion field.
    
\section{B. Modified Maxwell Equations in Axion Electrodynamics} \label{S3}  

Similarly, Maxwell equations can also be derived using a different method. To do this, first note that in 
classical electromagnetic theory, the fields $\vec{D}$ and $\vec{H}$ are written as
\begin{equation}
\vec{D} = \epsilon \vec{E} + \vec{P}
\end{equation}
\begin{equation}
\vec{H} = \vec{B} / \mu - \vec{M}\,. \label{HH}
\end{equation}
The expressions for $\vec{M}$ and $\vec{P}$ are derived from the Helmholtz free energy: 
$\vec{M} = -\partial \vec{F} / \partial \vec{B}$ and $\vec{P} = -\partial \vec{F} / \partial \vec{E}$,
\begin{equation}
\vec{D} = \epsilon \vec{E} - \beta \theta \vec{B} 
\end{equation}
\begin{equation}
\vec{H} = \vec{B} / \mu + \beta \theta \vec{E} \label{HH1}
\end{equation}
Here, $\epsilon$ is the dielectric tensor, $\mu$ is the magnetic permeability coefficient, and 
$\beta := 2\alpha / \pi Z_0$, with $\alpha := e^2 / 4\pi\epsilon_0 \hbar c$ is the fine structure 
constant, $Z_0 := \sqrt{\mu_0 /\epsilon_0}$ is the vacuum impedance, $e$ is the electron charge, and 
$c := 1/\sqrt{\epsilon_0 \mu_0}$ is the speed of light in vacuum. Hence, Maxwell equations change as 
follows:
\begin{equation}\label{nablaD}
\vec{\nabla} \cdot\left(\epsilon \vec{E}-\beta \theta \vec{B} \right)=\rho_f \Rightarrow \vec{\nabla} 
\cdot(\epsilon \vec{E})=\left(\rho_f+\rho_{\theta}\right)
\end{equation}
Here, $\rho_{\theta} = \beta \vec{\nabla} \theta \cdot \vec{B}$ is the axionic charge density. From 
Eq.~(\ref{HH}), following relations are obtained:
\begin{equation}\label{nablaD2}
\vec{\nabla} \times\left(\vec{B}/\mu + \beta \theta \vec{E} \right)=\vec{J}_f + 
\frac{\partial}{\partial t}\left(\epsilon \vec{E}-\beta \theta \vec{B} \right)
\end{equation}
\begin{equation}\label{eq:nablaB}
\vec{\nabla} \times \vec{B}=\mu \vec{J}_f+\epsilon \mu \frac{\partial \vec{E}}{\partial t} + 
\vec{J}_{\theta}\,,
\end{equation}
where, $\vec{J}_{\theta} = -\beta \mu (\dot{\theta} \vec{B} + \vec{\nabla} \theta \times \vec{E})$ is the 
current density due to the axion field (axion-induced current density). For a Dirac half-metal with 
boundaries at $0$ and $L$, $\vec{\nabla} \theta = \pi [\delta(z) \Theta(L-z) + \Theta(z) \delta(L-z)]$ 
and $\dot{\theta} = 0$. Using $\vec{H}_{\ell} = \frac{\vec{B}_{\ell}}{\mu}$, Eqs.~(\ref{nablaD}) and 
(\ref{nablaD2}) become:
\begin{equation}\label{b.B}
\vec{\nabla} \cdot \vec{D}_{\ell}=\rho_f+\rho_{\theta}
\end{equation}
\begin{equation}\label{b..B}
\vec{\nabla} \times \vec{H}_{\ell}=\partial_t \vec{D}_{\ell} + \vec{J}_f + \vec{J}_{\theta}
\end{equation}
The other two Maxwell equations remain unchanged:
\begin{equation}
\vec{\nabla} \cdot \vec{B}_{\ell}=0
\end{equation}
\begin{equation}
\vec{\nabla} \times \vec{E}_{\ell}+\partial_{t} \vec{B}_{\ell}=0 \label{eq317}
\end{equation}
These equations are Maxwell equations in the presence of the axion term, which we have been used in our 
calculations. In this study, in order to distinguish between the field values with and without the axion 
term, the subscript $\ell$ has been used in the fields when the axion is present. This letter signifies 
a linear field in the presence of the axion.

\section{C. Standard Boundary Conditions for a DSM Medium}

Once electromagnetic fields at each point in space are determined, boundary conditions at the boundaries 
of the DSM medium can also be found. To do this, a surface $\mathsf{S}$ is imagined that divides space 
into two separate regions. Standard boundary conditions for this surface are expressed as follows:

1) Tangential component of the electric field $\vec{E}$ at the interface is continuous: 
$\hat{n} \times(\vec{E}_1 - \vec{E}_2) = 0$.

2) Normal component of the magnetic field $\vec{B}$ perpendicular to the surface must be continuous: 
$\hat{n}\cdot(\vec{B}_1-\vec{B}_2) = 0$.

3) Component of the electric flux density vector $\vec{D}$ normal to the surface exhibits a 
``discontinuity" that depends on the surface charge density: $\hat{n}\cdot(\vec{D}_1-\vec{D}_2)=\rho^s$.

4) Tangential component of the $\vec{H}$ field to the surface exhibits a ``discontinuity" equivalent to 
the surface current density: $\hat{n}\times(\vec{H}_1-\vec{H}_2) = \vec{J}^s$.

Definition of $\hat{n}$ in the boundary conditions specified here represents the unit normal vector of 
the surface $\mathsf{S}$ directed from region `2' to region `1'. $\rho^s$ and $\vec{J}^s$ denote the 
surface charge and current densities, respectively. Standard boundary conditions for the DSM medium are 
obtained as shown in Table~\ref{table2}:

\begin{table}[!h]
\caption{Boundary conditions for a DSM of the wave configuration employed in TE mode, where 
$\tilde{\mathfrak{n}}_{\pm}$'s are the effective birefringence indices.}  \label{table2}
{\small
\begin{tabular}{@{\extracolsep{5pt}}ccl}
\toprule \hline
\\
$z=0$  & {~~~~~} &
$\begin{aligned}
&\mathbf{n}^2 \left[\left(A_{2+} + B_{2+}\right) + i \left(A_{2-} + B_{2-}\right)\right] = \left(1 + \frac{k_x}{\varepsilon_0 \omega} \right) \left(A_{1+} + B_{1+}\right) + i \left(1 - \frac{k_x}{\varepsilon_0 \omega} \right) \left(A_{1-} + B_{1-}\right)\\[3pt] \hline\\
&\left(1 - \frac{Z_0 \alpha}{\cos \theta}\right) \left[ A_{1+} -i A_{1-}\right] - \left(1 + \frac{Z_0 \alpha}{\cos \theta}\right) \left[ B_{1+} -i B_{1-}\right] = \mathbf{\tilde{n}}_+ \left(A_{2+} - B_{2+}\right) - i \mathbf{\tilde{n}}_- \left(A_{2-} - B_{2-}\right)\\[3pt]\hline\\
&\left(A_{2+} + B_{2+}\right) - i \left(A_{2-} + B_{2-}\right) = \left(A_{1+} + B_{1+}\right) - i \left(A_{1-} + B_{1-}\right)\\[3pt] \hline\\
&\left(A_{2+} + B_{2+}\right) + i \left(A_{2-} + B_{2-}\right) = \left(1 - \frac{Z_0 \sigma}{\sin \theta} \right) \left[ \left(A_{1+} + B_{1+}\right)  + i \left(A_{1-} + B_{1-} \right) \right]\\[3pt]  
\end{aligned}$\\
\hline \hline
&\\[-10pt]
$z=L$  & {~~~~~} &$\begin{aligned}
&\mathbf{n}^2\left[\left(A_{2+} e^{i k_z \mathbf{\tilde{n}}_+ L} + B_{2+} e^{-i k_z \mathbf{\tilde{n}}_+ L}\right) + i \left(A_{2-} e^{i k_z \tilde{n}_- L} + B_{2-} e^{-i k_z \tilde{n}_- L}\right)\right] \notag \\
&= \left(1 - \frac{k_x}{\varepsilon_0 \omega} \right) \left(A_{3+} e^{i k_z L} + B_{3+} e^{-i k_z L}\right) + i \left(1 + \frac{k_x}{\varepsilon_0 \omega} \right) \left(A_{3-} e^{i k_z L} + B_{3-} e^{-i k_z L}\right)\\[3pt] \hline\\
&\mathbf{\tilde{n}}_+ \left[A_{2+} e^{i k_z \mathbf{\tilde{n}}_+ L} - B_{2+} e^{-i k_z \mathbf{\tilde{n}}_+ L}\right] - i \mathbf{\tilde{n}}_- \left[A_{2-} e^{i k_z \mathbf{\tilde{n}}_- L} - B_{2-} e^{-i k_z \mathbf{\tilde{n}}_- L}\right]  \notag \\
&= \left(1 + \frac{Z_0 \alpha}{\cos \theta} \right) \left[A_{3+} - i A_{3-} \right] e^{i k_z L}  - \left(1 - \frac{Z_0 \alpha}{\cos \theta} \right) \left[B_{3+} - i B_{3-}\right] e^{-i k_z L}\\[3pt] \hline\\  
&\left(A_{2+} e^{i k_z \mathbf{\tilde{n}}_+ L} + B_{2+} e^{-i k_z \mathbf{\tilde{n}}_+ L}\right) - i \left(A_{2-} e^{i k_z \mathbf{\tilde{n}}_- L} + B_{2-} e^{-i k_z \mathbf{\tilde{n}}_- L}\right) = \left(A_{3+} -i A_{3-} \right) e^{i k_z L} + \left(B_{3+} - i B_{3-} \right) e^{-i k_z L}\\[3pt]  \hline\\ 
&\left[A_{2+} e^{i k_z \mathbf{\tilde{n}}_+ L} + B_{2+} e^{-i k_z \mathbf{\tilde{n}}_+ L}\right] + i \left[A_{2-} e^{i k_z \mathbf{\tilde{n}}_- L} - B_{2-} e^{-i k_z \mathbf{\tilde{n}}_- L}\right] \notag \\
&= \left(1 + \frac{Z_0 \sigma}{\sin \theta} \right) \left[ \left( A_{3+} + i A_{3-}\right)  e^{i k_z L} + \left(B_{3+} + i B_{3-} \right) e^{-i k_z L} \right]
\end{aligned}$\\[-8pt]
&\\
\hline \hline
\end{tabular}
}
\end{table}

\section*{Acknowledgement} This work is funded by Scientific Research Projects Coordination Unit (BAP) of 
Istanbul University Project Number FBA-2020-35018.

\textbf{Competing Interests:} The authors declare no competing interests. 

\textbf{Data Availability Statement:} The authors declare that any data that support the findings of this 
study are included within the article.

\end{document}